\documentclass{article}
\usepackage[right=2.5cm,left=2.5cm,top=2.25cm,bottom=2.5cm,headsep=0.5cm,footskip=1cm]{geometry}
\usepackage[utf8]{inputenc}
\usepackage[spanish]{babel}
\usepackage{graphicx}
\usepackage{subfigure}
\usepackage{amsmath, amsthm, amsfonts}
\usepackage{multicol}
\usepackage{float}
\graphicspath{ {c:/ima/} }
\usepackage{fancyhdr}
\fancypagestyle{plain}{
\vspace{0.3cm}

\fancyfoot[L]{\textsuperscript{1}hylasernad@correo.udistrital.edu.co}
\fancyfoot[C]{\textsuperscript{2}slsilvas@correo.udistrital.edu.co}
\fancyfoot[R]{\textsuperscript{3}ycruz@pedagogica.edu.co}
}

\begin{document}

\title {\LARGE\textbf{SIMULATION ON VARIOUS TOPOCENTRIC POINTS OF THE ZODIAC CONSTELLATIONS\\ \large{Simulación en Diversos Puntos Topocéntricos de las Constelaciones Zodiacales}}}

\author{\Large{H. Laserna\textsuperscript{1}, S. Silva\textsuperscript{2}, Y. Cruz\textsuperscript{3}}\\
\normalsize \textsuperscript{1,2,3}Grupo de F\'isica e Inform\'atica FISINFOR, Universidad Distrital Francisco Jos\'e de Caldas,\\ Facultad de Ciencias y Educaci\'on, Licenciatura en F\'isica.}
\date{Entregado 27/12/2017}
\maketitle \textbf{Abstract. }On the sighting of the constellations and their stars, they seem to be near each other at first sight but some of them are actually quite further from the others. From the geocentric point of view can the Zodiac Constellations be perceived as referential marks divided into the terrestrial ecliptic, which are organized according to their shapes.In this paper are these star groups simulated in order to compare the shapes of these groups and the diverse coordinate systems of the astronomy. Within the simulation is the distance measure light-year replaced with a so-called program's default “unity" that uses a compilation in Python,where such distances are escalated to 1:10.
\\

\maketitle \textbf{Resumen. }Al observar las constelaciones y sus estrellas, a simple vista parecen estar cercanas entre sí, cuando en realidad unas están mucho más lejanas que otras. Desde el punto de vista geocéntrico se perciben las constelaciones zodiacales como marcas de referencia divididas en la eclíptica terrestre en donde son organizadas según sus formas. En este trabajo se simulan éstas agrupaciones de astros para comparar desde distintos puntos topocéntricos la forma de las mismas y los distintos sistemas de coordenadas en astronomía. Dentro de la simulación se remplaza la medida de distancia de años luz por una llamada “unidad" predeterminada del programa que utiliza una compilación en Python, donde se escalan dichas distancias a 1:10.\\
\\
\small {\textbf{Palabras Clave:} Constelaciones Zodiacales, Simulación, Astronomía, Coordenadas, Eclíptica.}
\\
\begin{multicols}{2}
\section{The Celestial Sphere}
When one observes the stars shining in the firmament of a beautiful clear night, we notice differences in their brightness and in their apparent arrangements that allows us to even think of associating figures with such groups. We call these arrangements constellations; We currently count 88 constellations. In the perception of the direct visualization of the night sky, one does not conceive the distances to the stars, then the notion of an apparently spherical surface appears where the stars seem to be fixed to it. It is this apparent perception that led ancient astronomers to the concept of the celestial sphere.\\
\\
Today we maintain the concept of celestial sphere but not as the spherical surface where the stars are because that notion is wrong; nevertheless, we maintain a geometric concept that allows us to describe very well the positions of the stars in the sky, by means of the construction of astronomical coordinate systems, and by means of the suitable use of this tool we can locate very well any celestial body and thus make observations astronomical.\\
\\
For us, the celestial sphere is a sphere of unitary radius, such that any scale of distance in the solar system turns out to be totally imperceptible in front of the radius of the celestial sphere. The celestial sphere is an apparent surface of the heavens at infinite distance from Earth and on the surface from which the stars appear as fixed. However, the fact of having an immensely large unit radius compared to distances in the solar system has practical effects.\\
\\
The celestial sphere is classified according to the point where the center of the sphere is defined conceptually in the following way:

\begin{itemize}
\item Topocentric: centered on the observer
\item Geocentric: centered in the center of the Earth
\item Heliocentric: centered in the center of the Sun
\item Baricentric: centered on the center of gravity of a system
\end{itemize}

Thus, a celestial sphere where the visualization of the cardinal points and the angle of "elevation" of one of the celestial poles for a given observer predominates, is a celestial sphere that makes a lot of sense to locate it with a center in it, the observer, taking into account both a topocentric celestial sphere.

\section{Horizontal coordinates}
Given the horizontal plane of the observer, we extend it indefinitely and intersect the celestial sphere to obtain a maximum circle called celestial horizon, which divides the celestial sphere into two hemispheres (like any maximum circle), one that at that moment is visible to the observer and the other who cannot observe it.\\
\\
The horizontal astronomical coordinates are very important and derive their name because the maximum fundamental circle of these coordinates is the celestial horizon. The associated concepts are known: celestial horizon, nadir, zenith and the vertical ones. But it remains to define a point of reference on the celestial horizon and a sense. To do this, the cardinal points are defined.\\
\\
The cardinal point North (N) is the point that results from the intersection of the celestial horizon with the vertical that passes through the north celestial pole. The South cardinal point (S) is the point of intersection of the celestial horizon with the vertical that passes through the celestial south pole. The intersection of the celestial horizon with the equator generates two antipodal points, the cardinal points East (E) and West (W) with the western cardinal point being on the side of the apparent diurnal movement of the stars.\\
\\
Then, given a star, its azimuth and height horizontal coordinates are defined as follows:

\begin{itemize}
\item We trace the vertical of the star.
\item We will call azimuth to the length of the arc drawn on the celestial horizon, from the North cardinal point, following the East direction, to the vertical of the star.
\item The height of the star is the length of the arc drawn on the star's vertical, from the celestial horizon to the star.
\end{itemize}

So: 0\textsuperscript{$\circ$} $\leq$ A $<$ 360\textsuperscript{$\circ$} ; -90\textsuperscript{$\circ$} $\leq$ h $\leq$ 90\textsuperscript{$\circ$}. Where A denotes the azimuth of the star and h its height.

\includegraphics[scale=0.7]{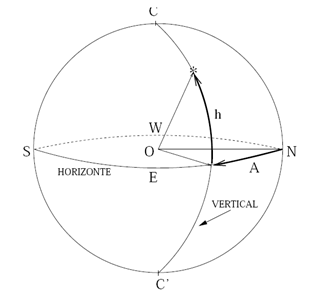}\\

Figure 1. Horizontal coordinates. Source: Portilla, G. (2001).

\section{Orthogonal Projection Systems in Technical Drawing and Views}
All representation systems have the objective of representing on a two-dimensional surface, such as a sheet of paper, objects that are three-dimensional in space.\\
\\
All systems are based on the projection of objects on a plane, which is called the plane of the frame or projection, by means of the so-called projecting rays. The number of projection planes used, the relative situation of these with respect to the object, as well as the direction of the projecting rays, are the characteristics that differentiate the different representation systems.\\
\\
In all systems of representation, the projection of objects on the plane of the picture or projection, is done by the projecting rays, these are imaginary lines, passing through the vertices or points of the object, provide at its intersection with the plane of the painting, the projection of said vertex or point.\\
\\
They are called main views of an object, orthogonal projections of the same on 6 planes, arranged in cube form. You could also define the views as, the orthogonal projections of an object, according to the different directions from which you look.\\
\\
If we place an observer according to the six directions indicated by the arrows, we would obtain the six possible views of an object.\\
\\
These views are given the following names:

\includegraphics[scale=0.9]{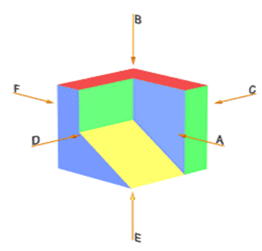}\\

Figure 2. Views of an Object. Source:www.larapedia.com.

\begin{itemize}
\item View A: Front or up view
\item View B: Top view or floor
\item View C: Right or right side view
\item View D: Left or left side view
\item View E: Bottom view
\item View F: Rear view
\end{itemize}

\section{Simulations}
Taking into account the previous concepts, the Zodiacal Constellations are modeled and observed according to the orthogonal projection system ASA, in the frontal and right lateral perspectives where the distances of the stars will be observed through the projecting rays.\\

\begin{itemize}
\item Analyze the topocentric positions of the Constellations to demonstrate the relationships of their celestial bodies.
\item Use simulations as a means of teaching-learning of physical models in astronomy.
\end{itemize}

It is arranged to simulate the zodiacal constellations by means of a data base of the distances of the earth to each star, the azimuth and the length for later with said data to enter it in Autodesk Maya by means of compilation in python, then to configure a camera that bar 90\textsuperscript{$\circ$} by the horizontal with a constant radius and then render that image when the camera is in a right lateral position.

\subsection{Pisces constellation}
\includegraphics[scale=0.1]{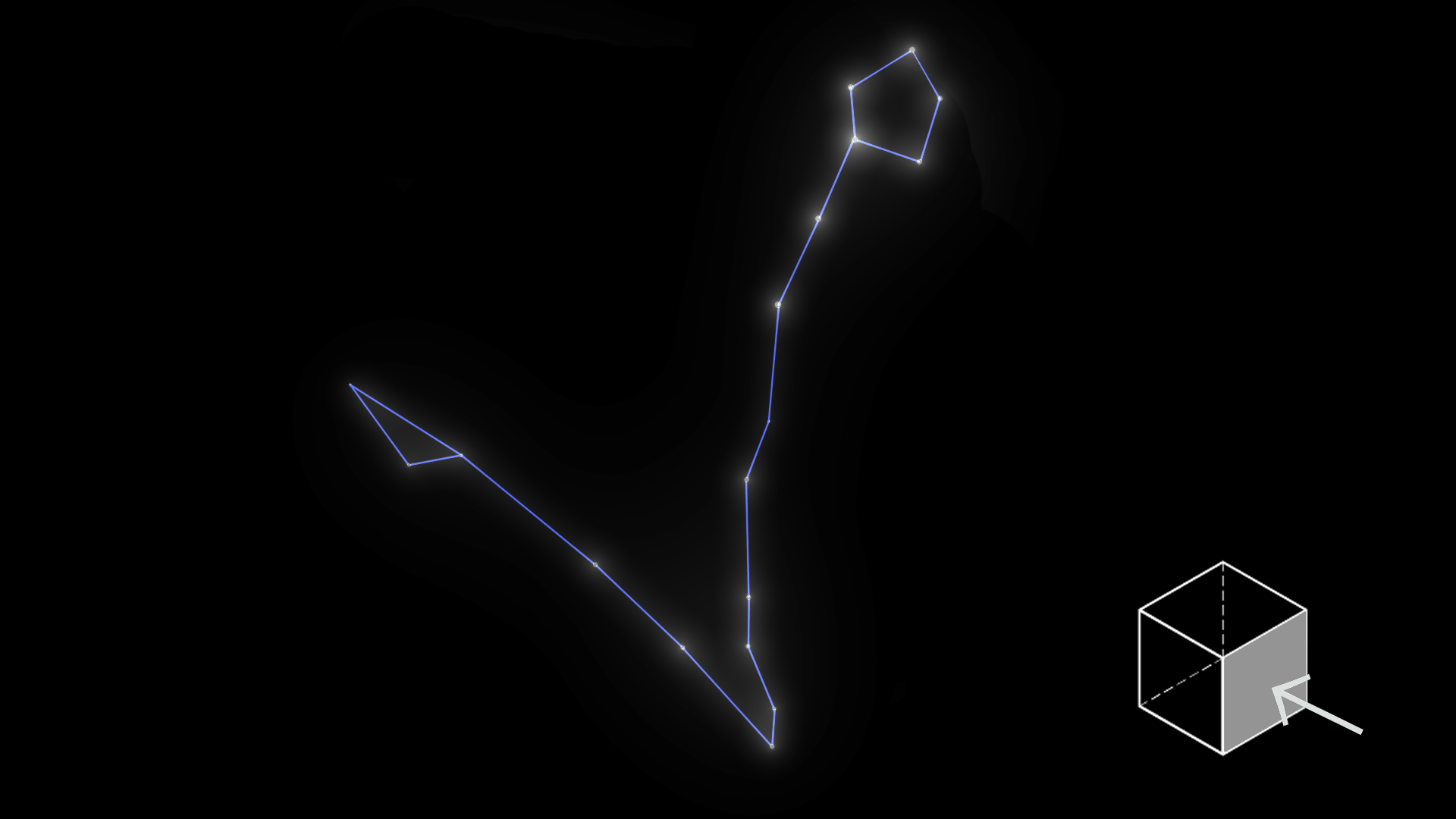}\\
Figure 3. Pisces-Front view. Author: Laserna, H. (2016).\\

\includegraphics[scale=0.1]{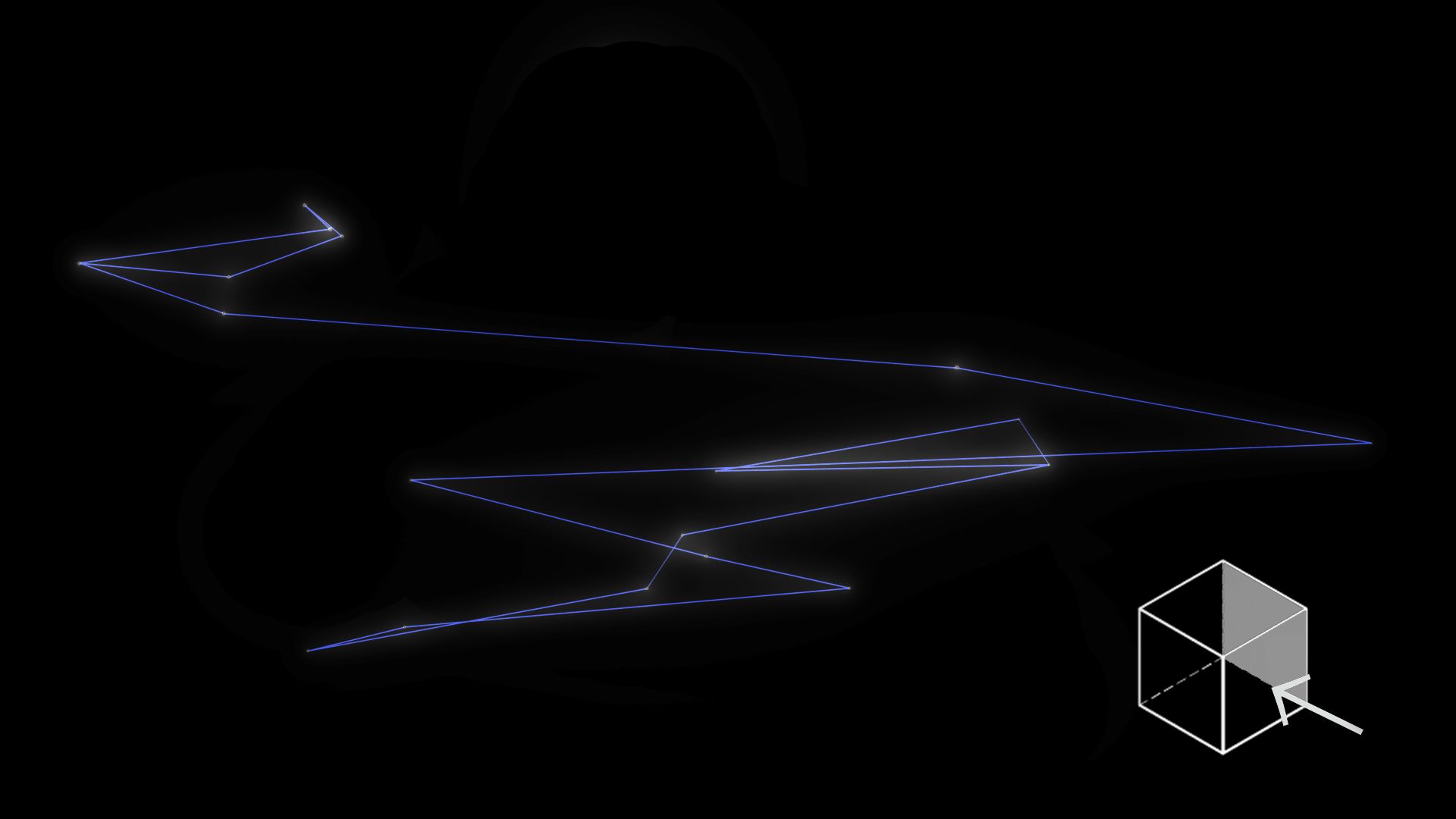}\\
Figure 4. Pisces-Left view. Author: Laserna, H. (2016).\\

When we change our view to the left side we can perceive the separation of the stars that make it up.

\subsection{Aries constellation}
\includegraphics[scale=0.1]{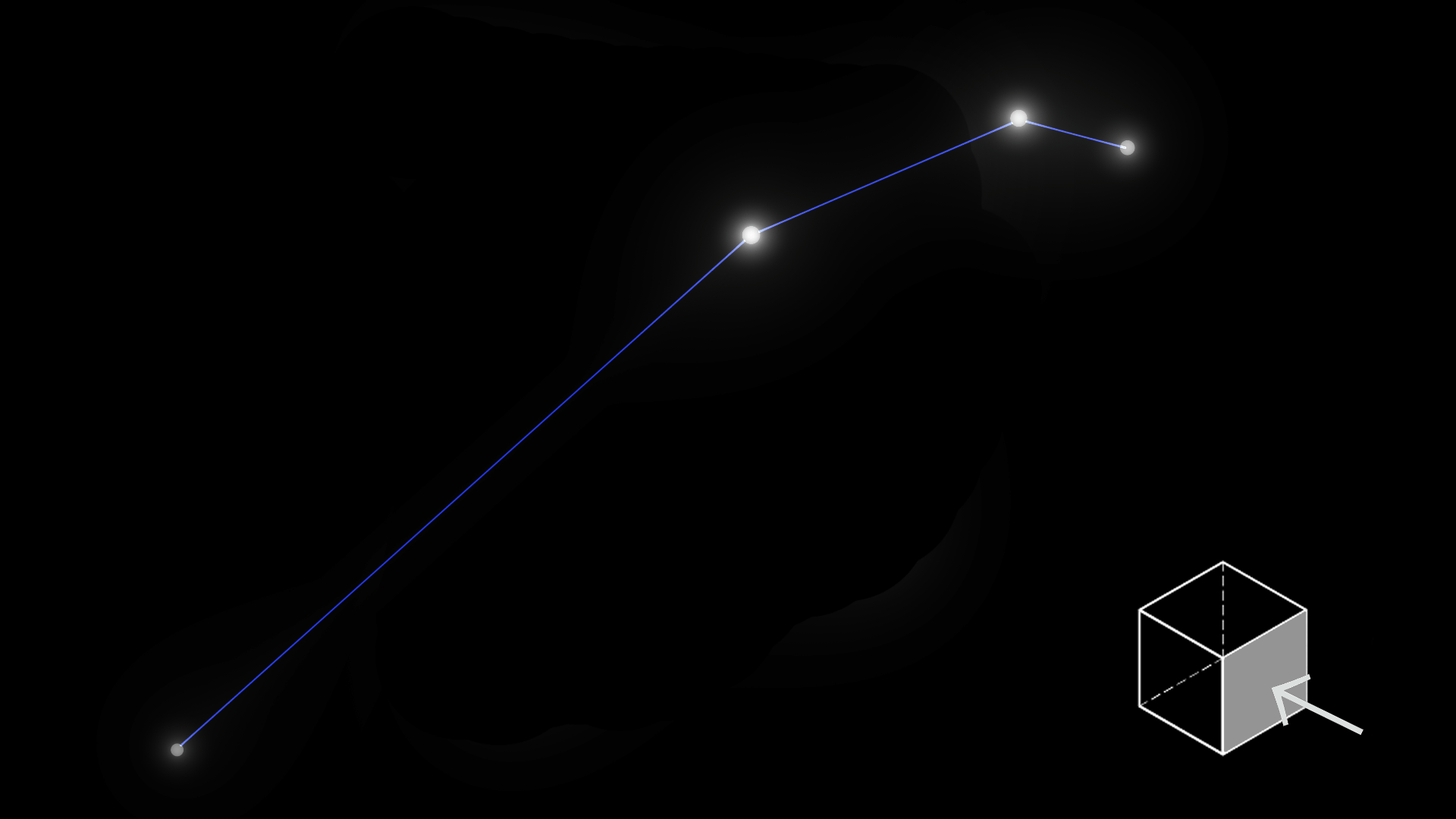}\\
Figure 5. Aries-Front view. Author: Laserna, H. (2016).\\

\includegraphics[scale=0.1]{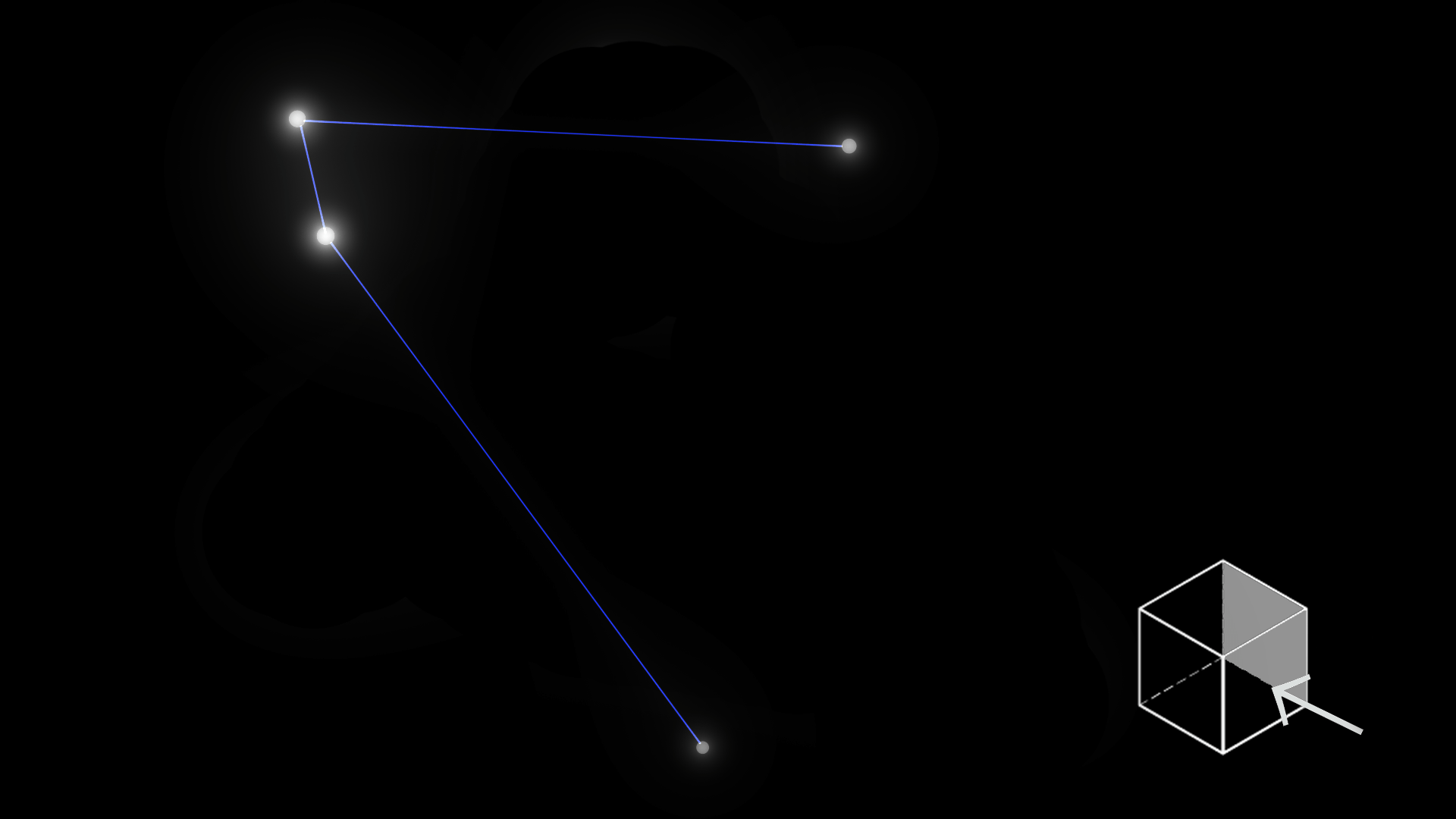}\\
Figure 6. Aries-Left view. Author: Laserna, H. (2016).\\

\subsection{Taurus constellation}
\includegraphics[scale=0.1]{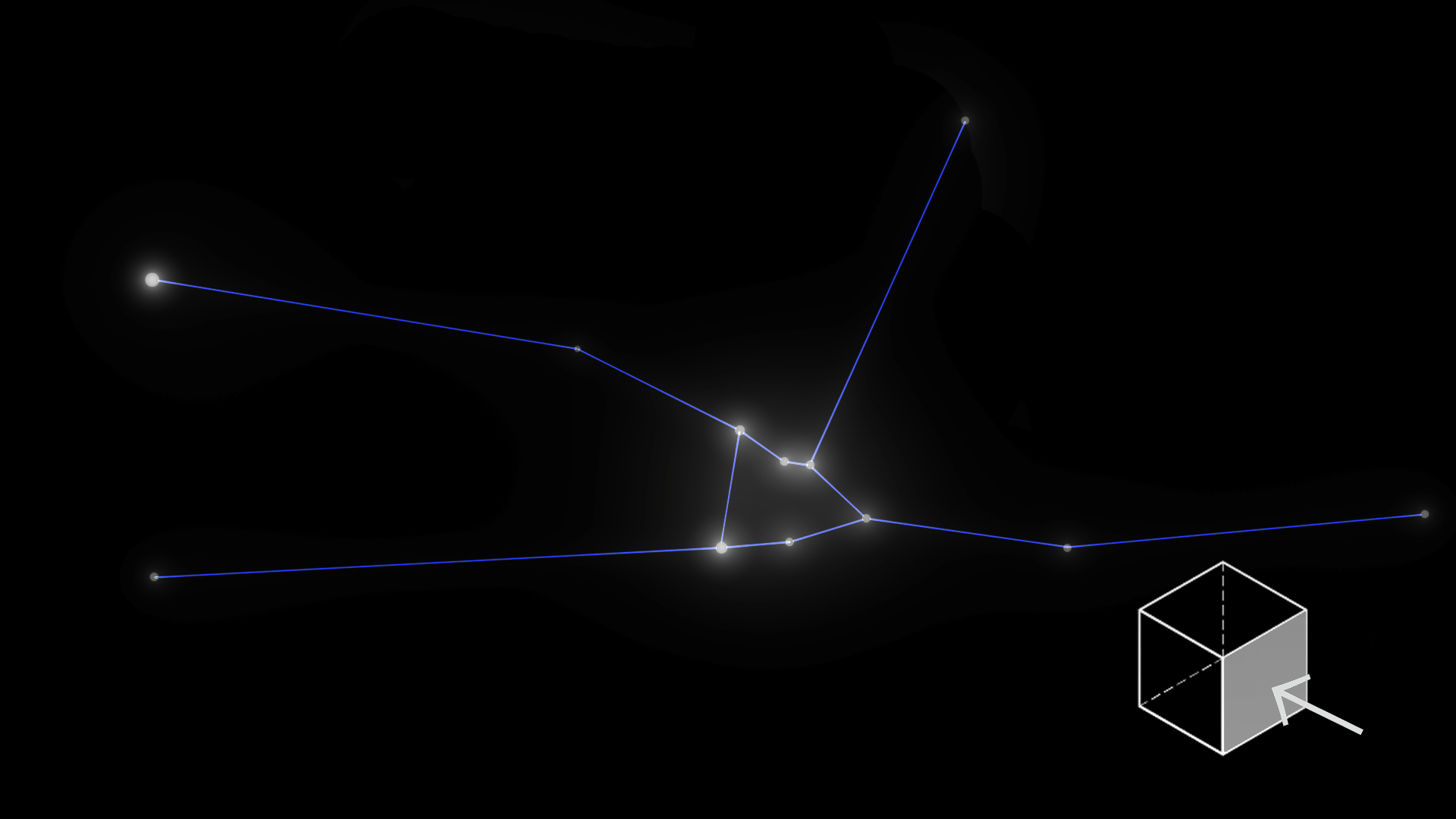}\\
Figure 7. Taurus-Front view. Author: Laserna, H. (2016).\\

\includegraphics[scale=0.1]{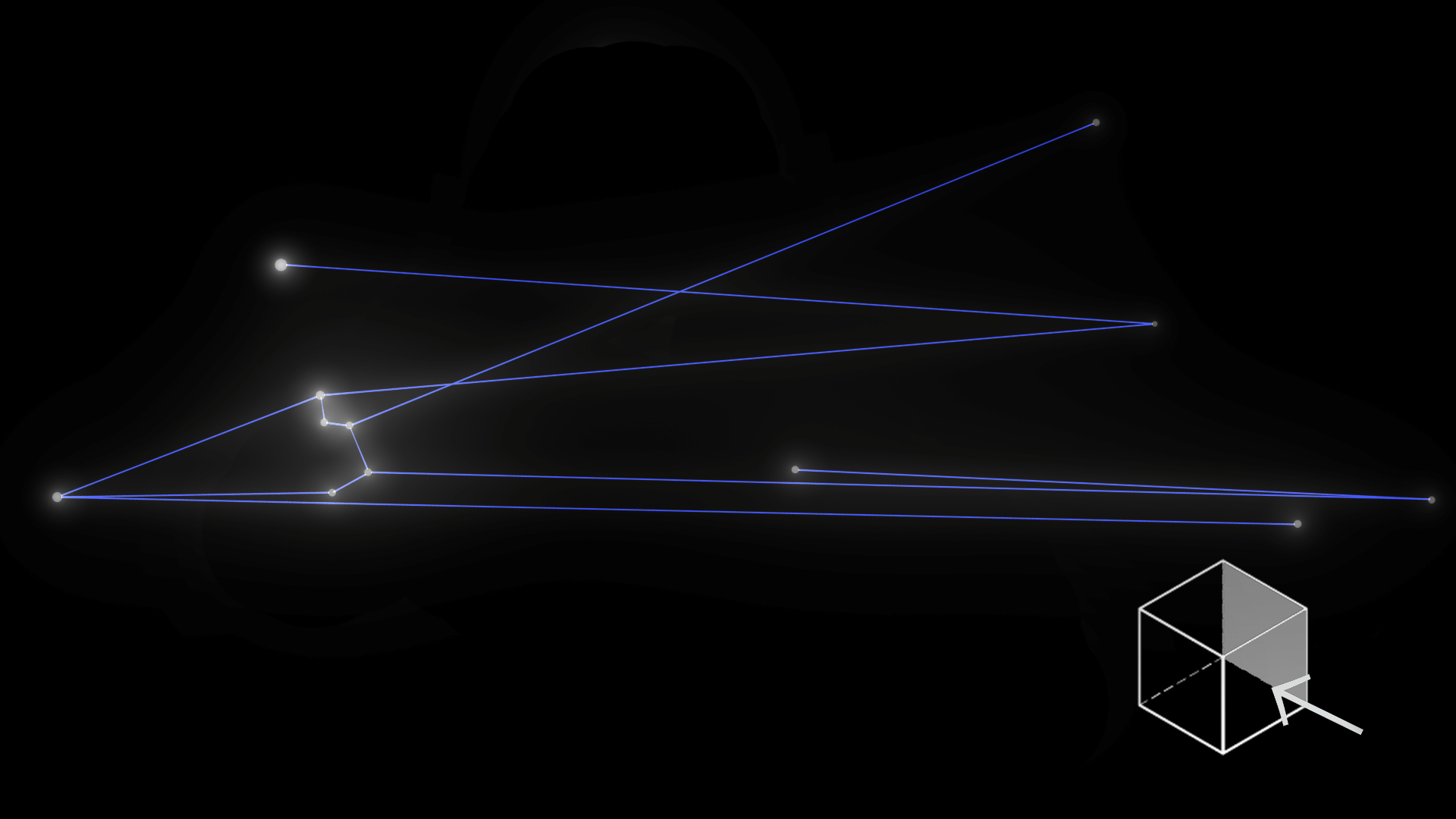}\\
Figure 8. Taurus-Left view. Author: Laserna, H. (2016).\\

\subsection{Gemini constellation}
\includegraphics[scale=0.1]{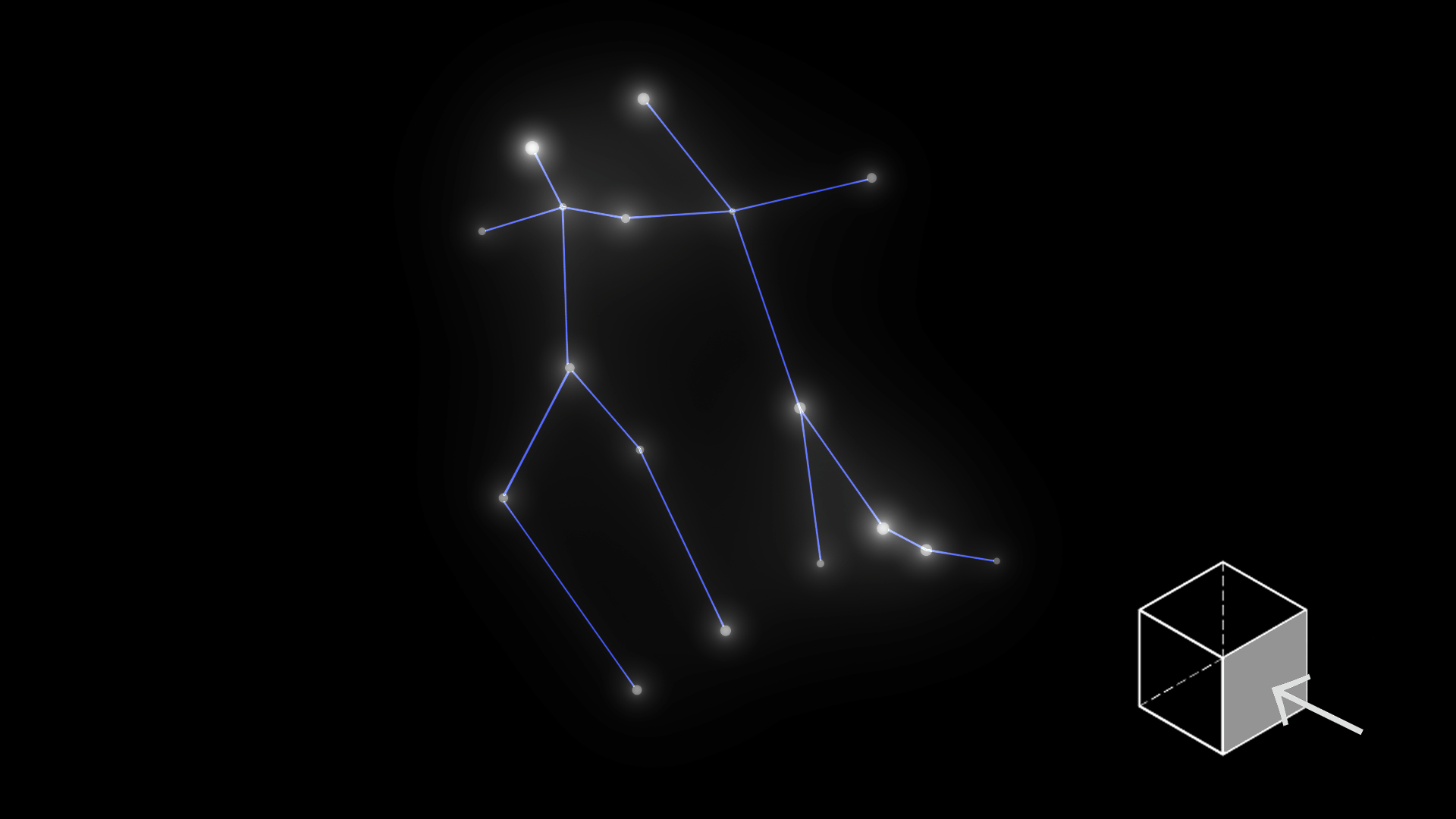}\\
Figure 9. Gemini-Front view. Author: Laserna, H. (2016).\\

\includegraphics[scale=0.1]{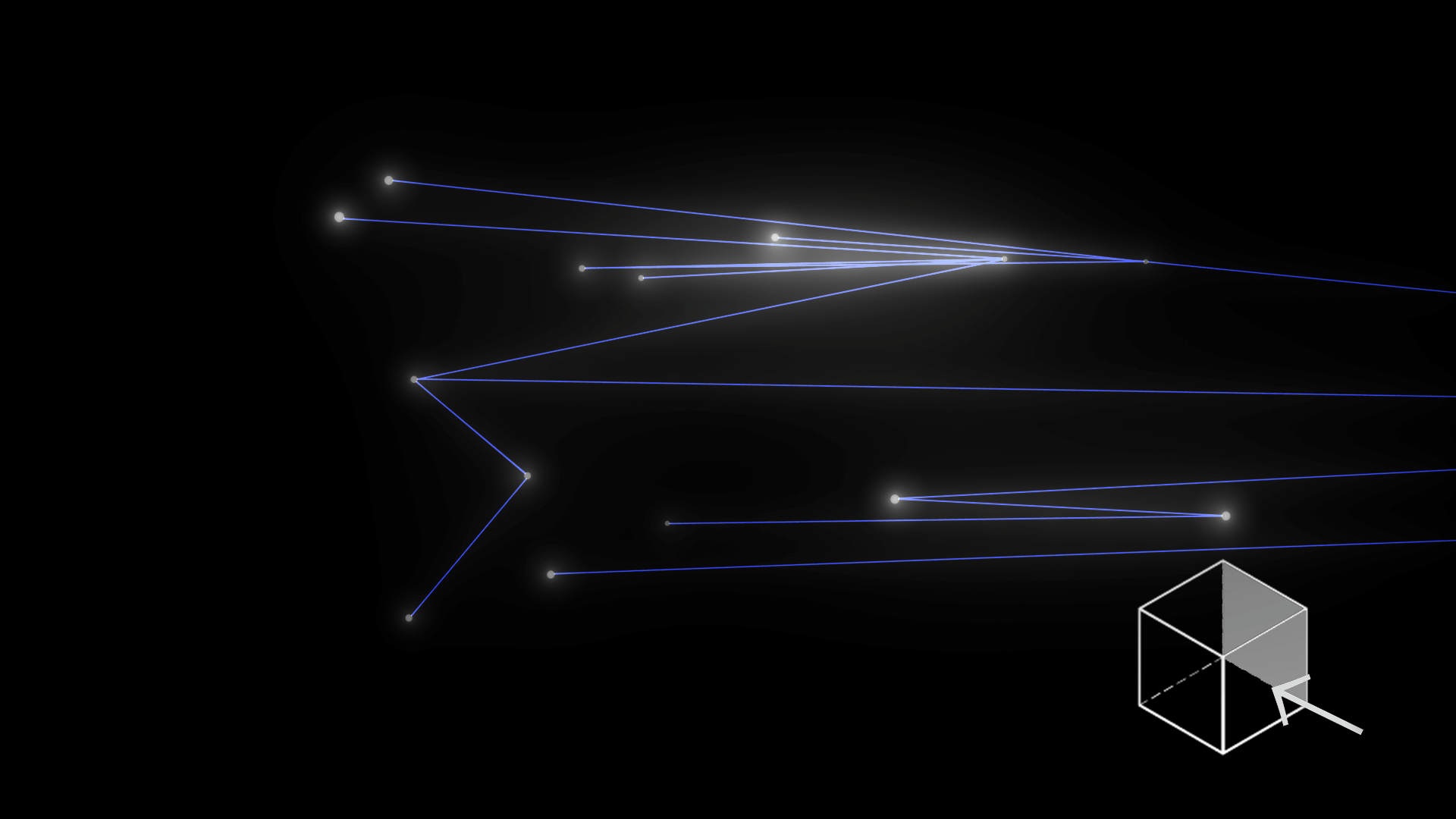}\\
Figure 10. Gemini-Left view. Author: Laserna, H. (2016).\\

\subsection{Cancer constellation}
\includegraphics[scale=0.1]{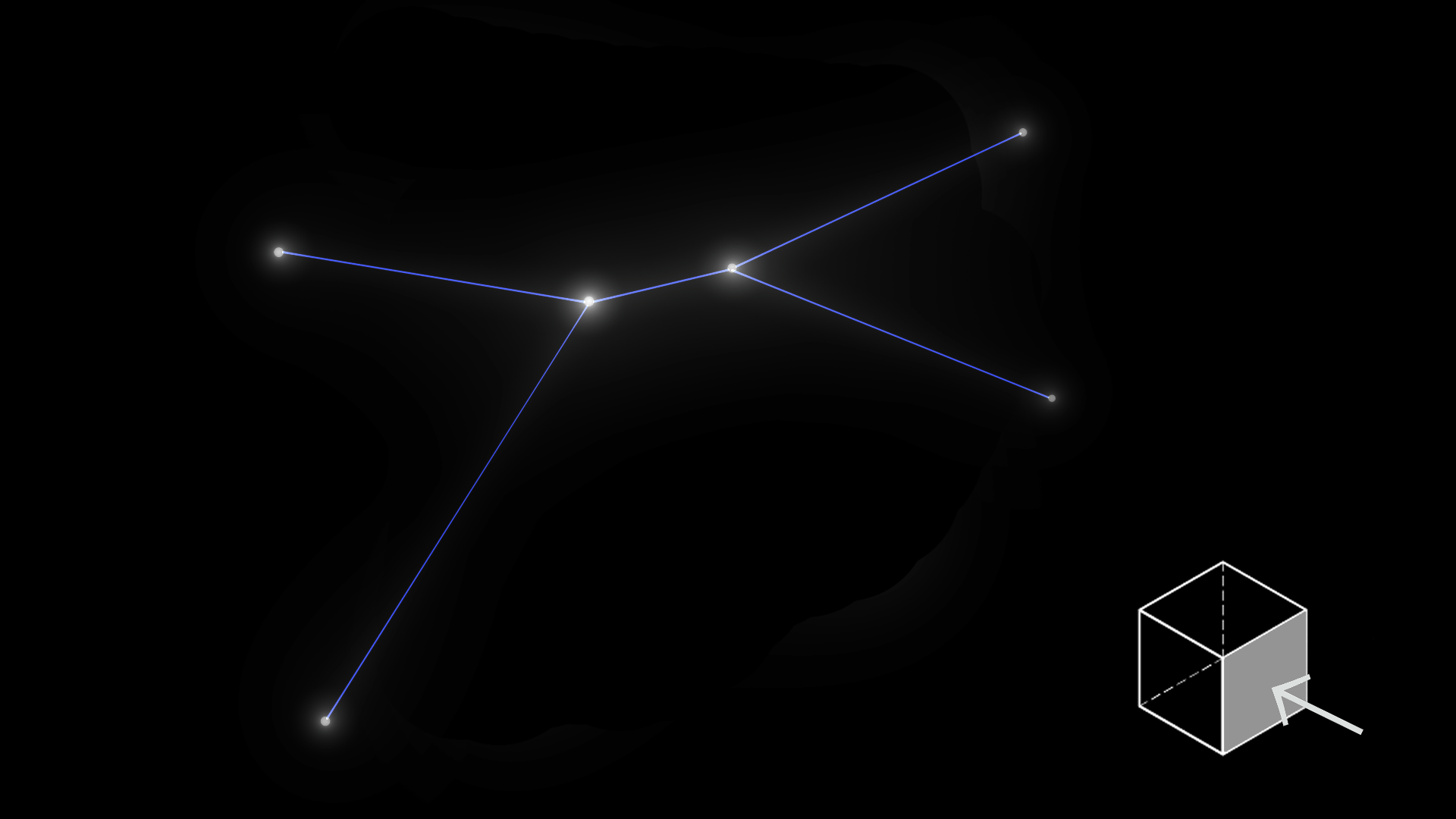}\\
Figure 11. Cancer-Front view. Author: Laserna, H. (2016).\\

\includegraphics[scale=0.1]{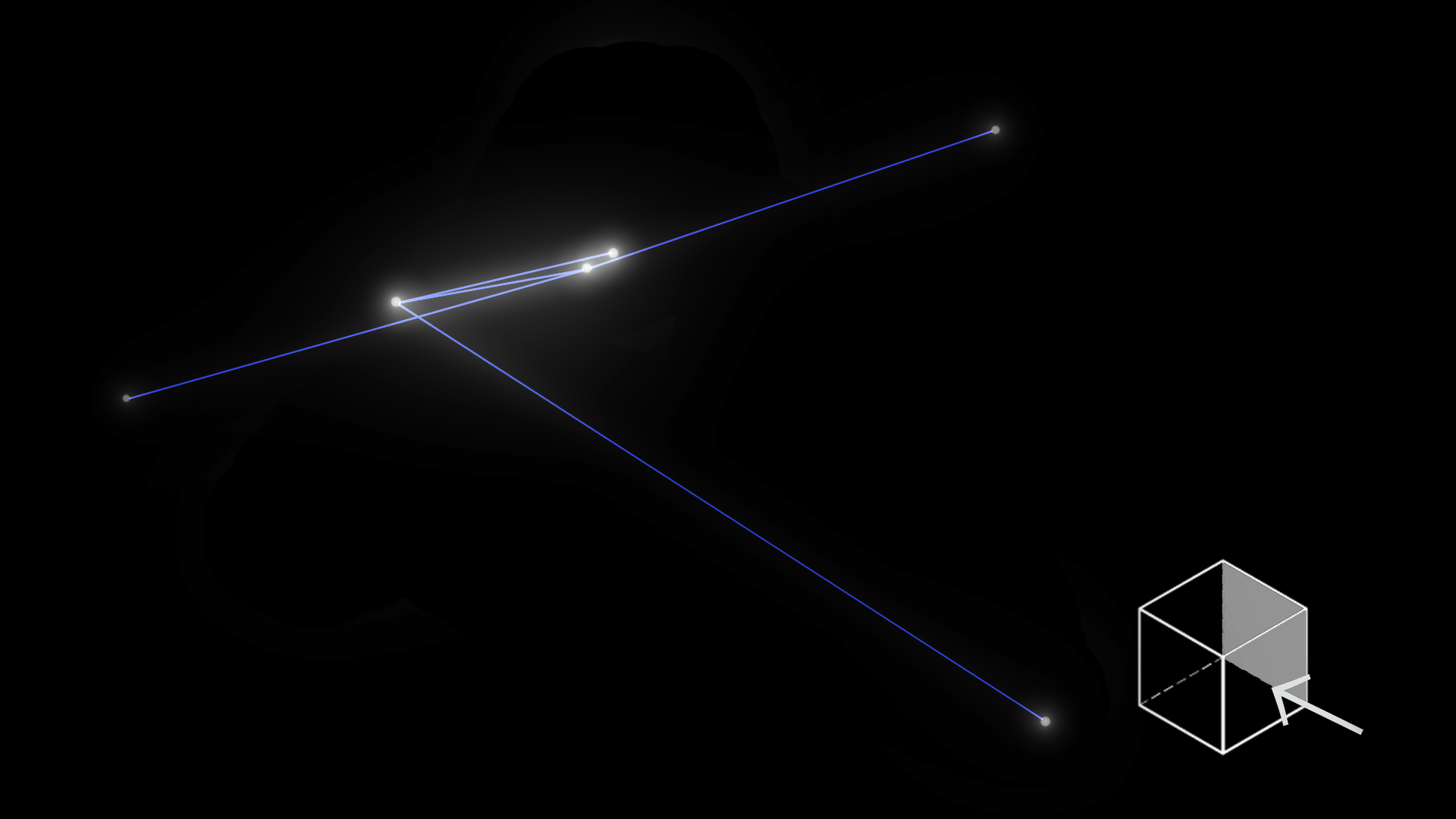}\\
Figure 12. Cancer-Left view. Author: Laserna, H. (2016).\\

\subsection{Leo constellation}
\includegraphics[scale=0.1]{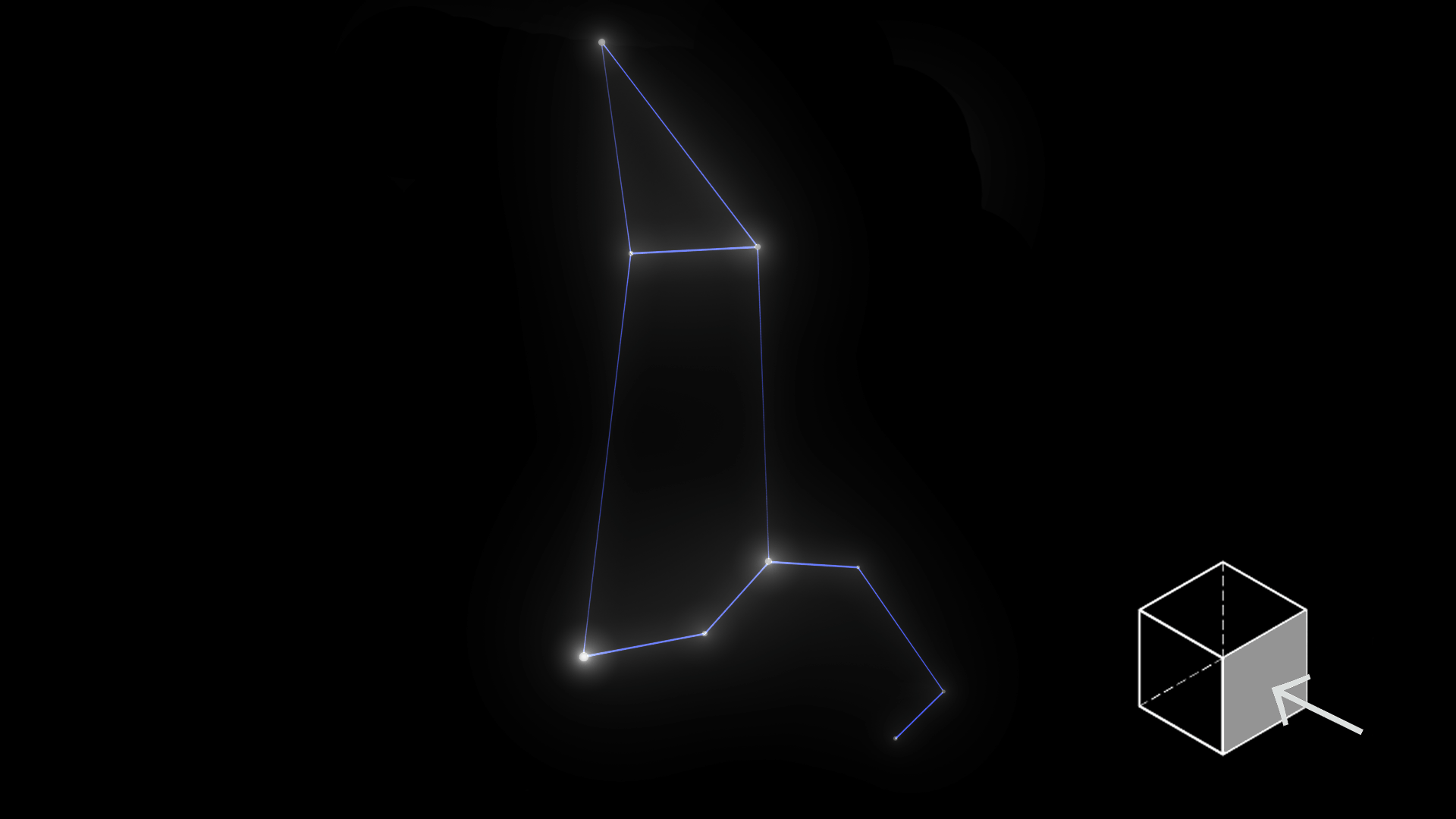}\\
Figure 13. Leo-Front view. Author: Laserna, H. (2016).\\

\includegraphics[scale=0.1]{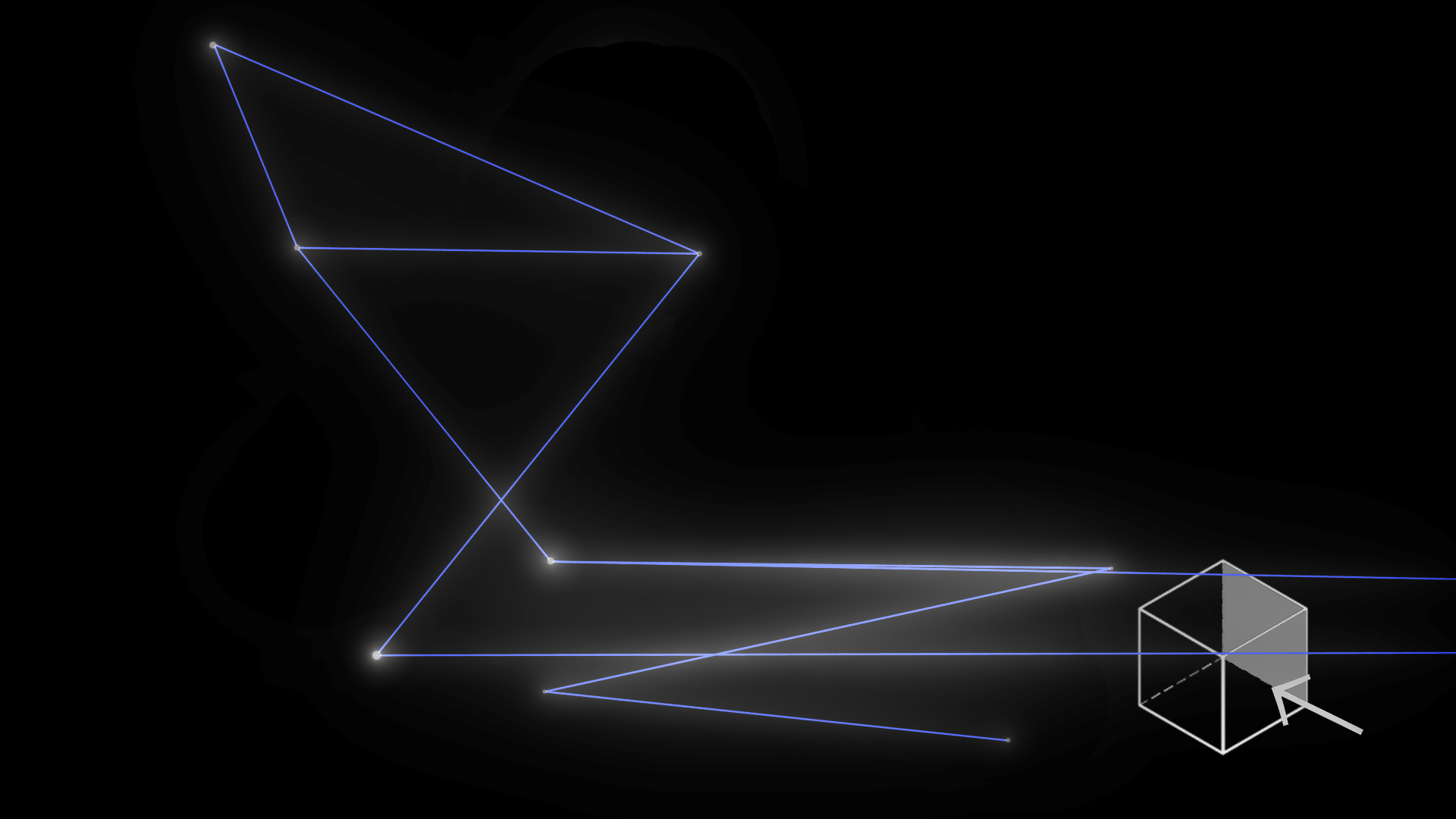}\\
Figure 14. Leo-Left view. Author: Laserna, H. (2016).\\

\subsection{Virgo constellation}
\includegraphics[scale=0.1]{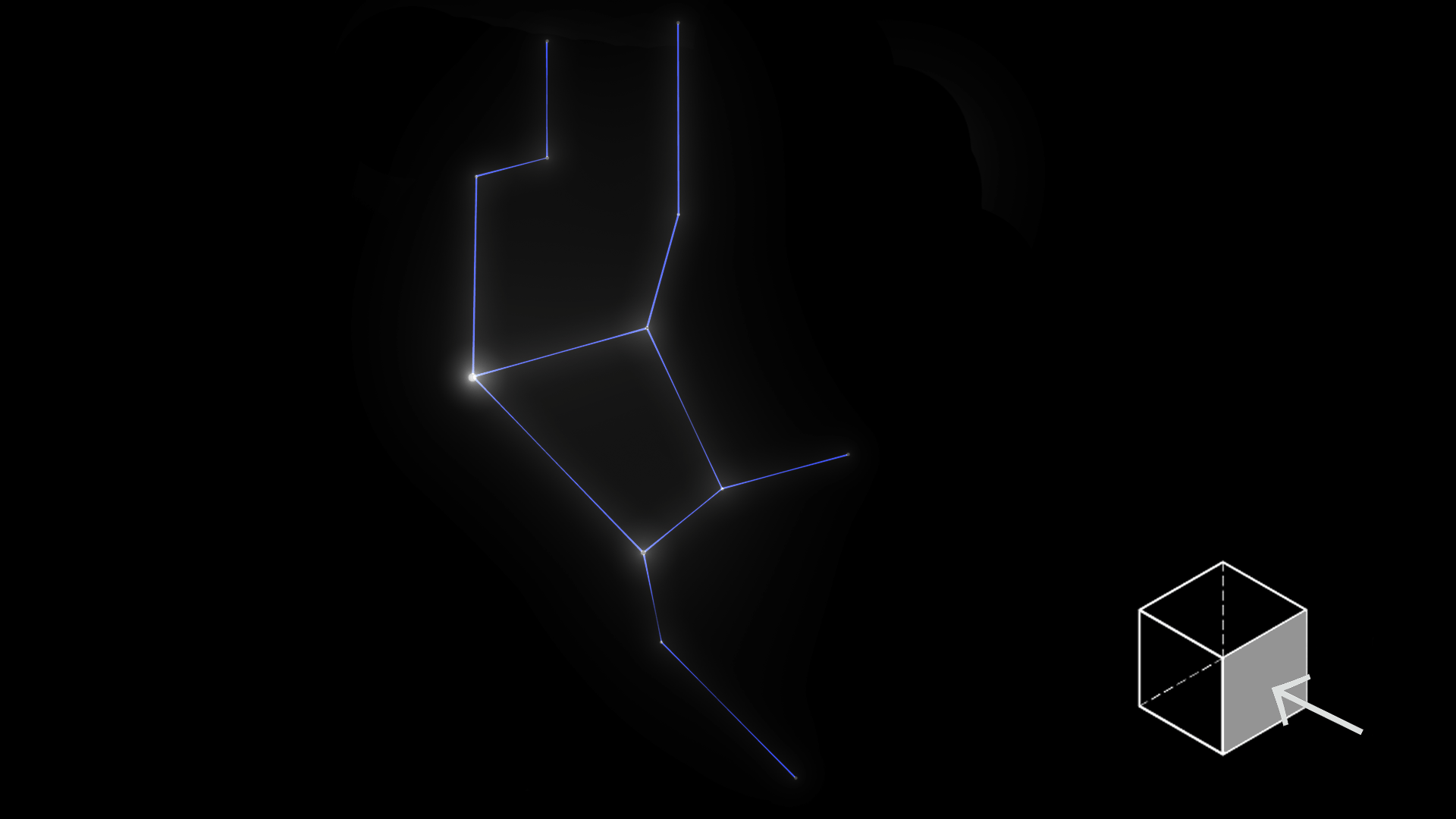}\\
Figure 15. Virgo-Front view. Author: Laserna, H. (2016).\\

\includegraphics[scale=0.1]{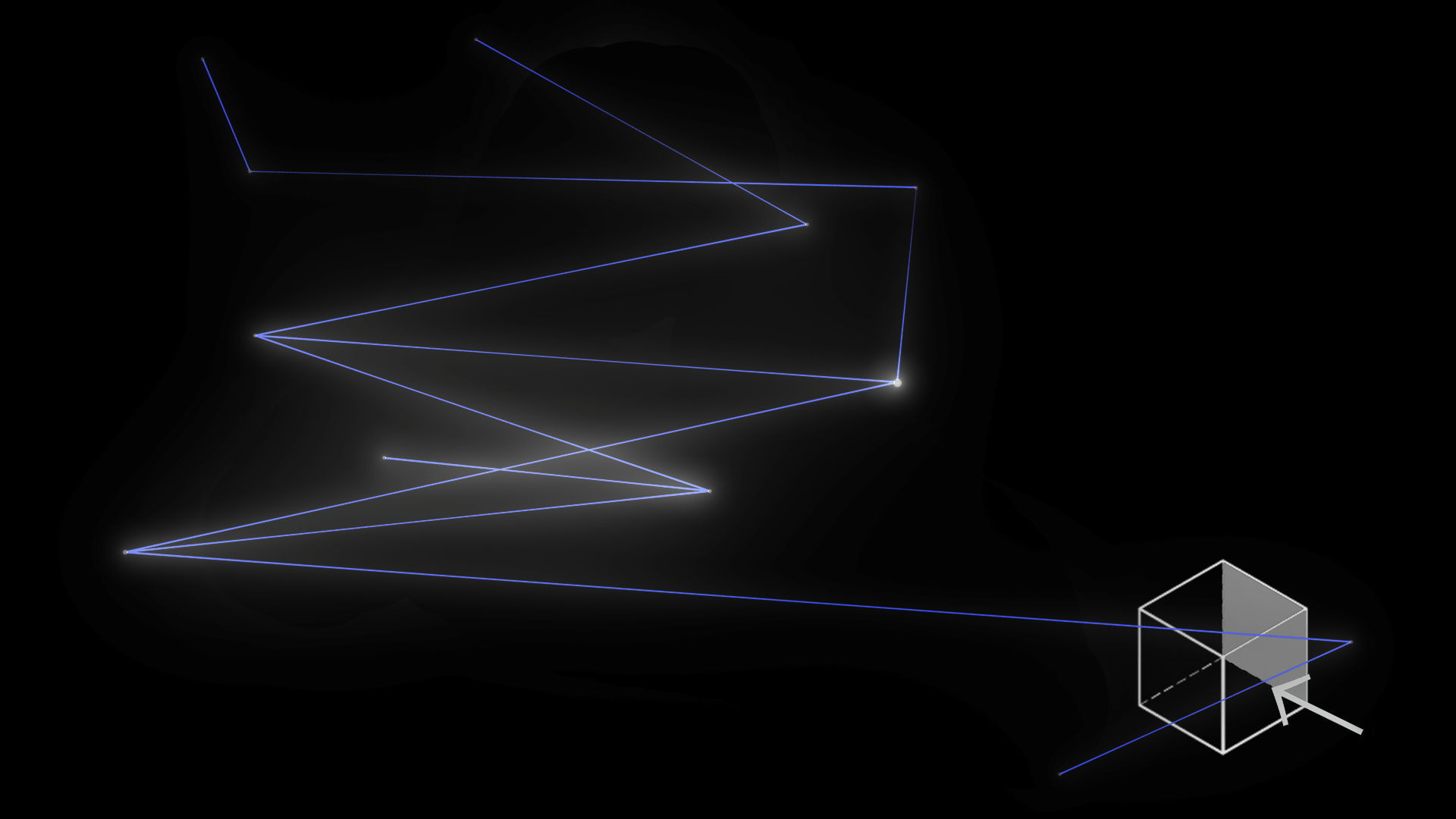}\\
Figure 16. Virgo-Left view. Author: Laserna, H. (2016).\\

\subsection{Libra constellation}
\includegraphics[scale=0.1]{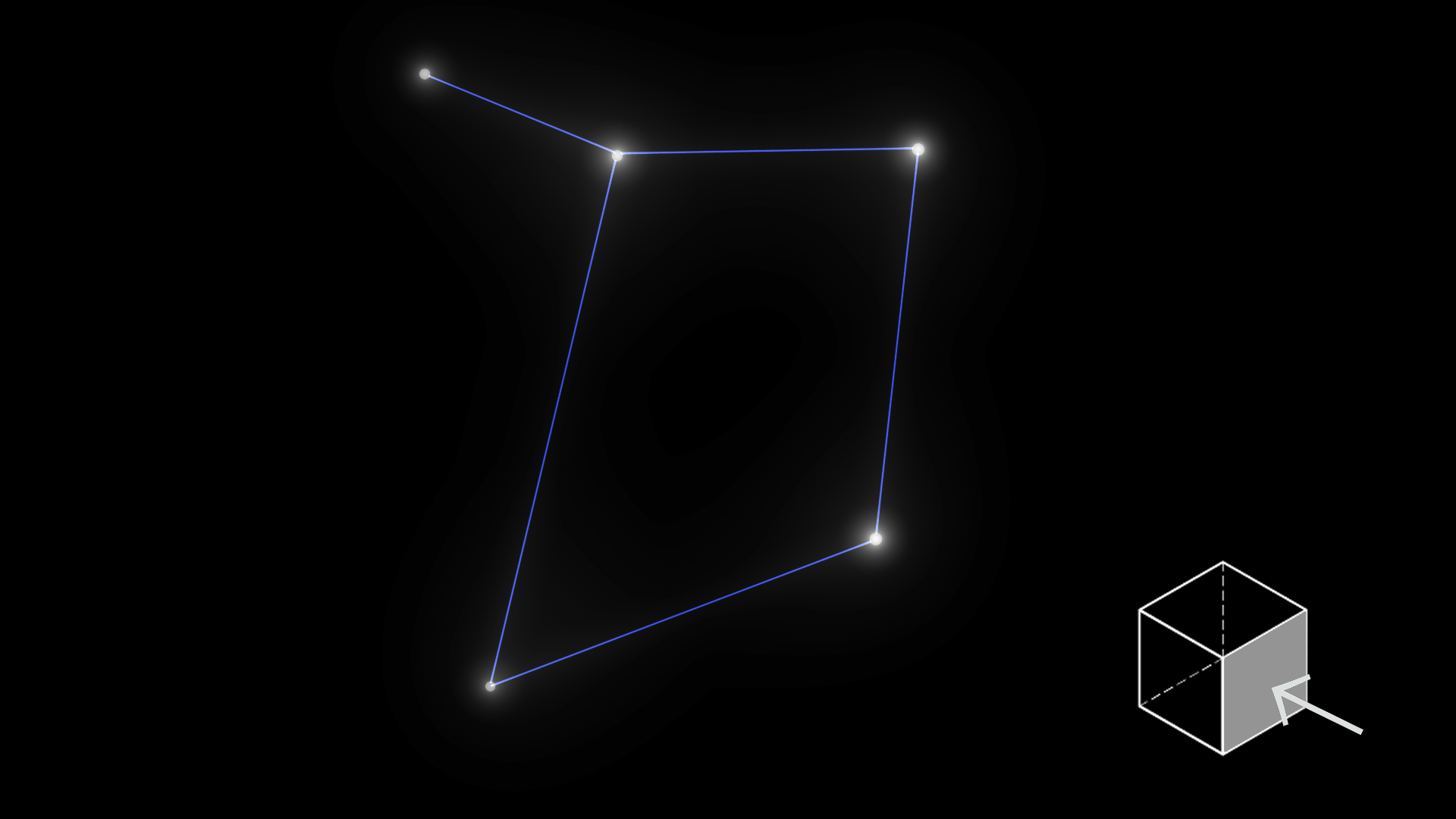}\\
Figure 17. Libra-Front view. Author: Laserna, H. (2016).\\

\includegraphics[scale=0.1]{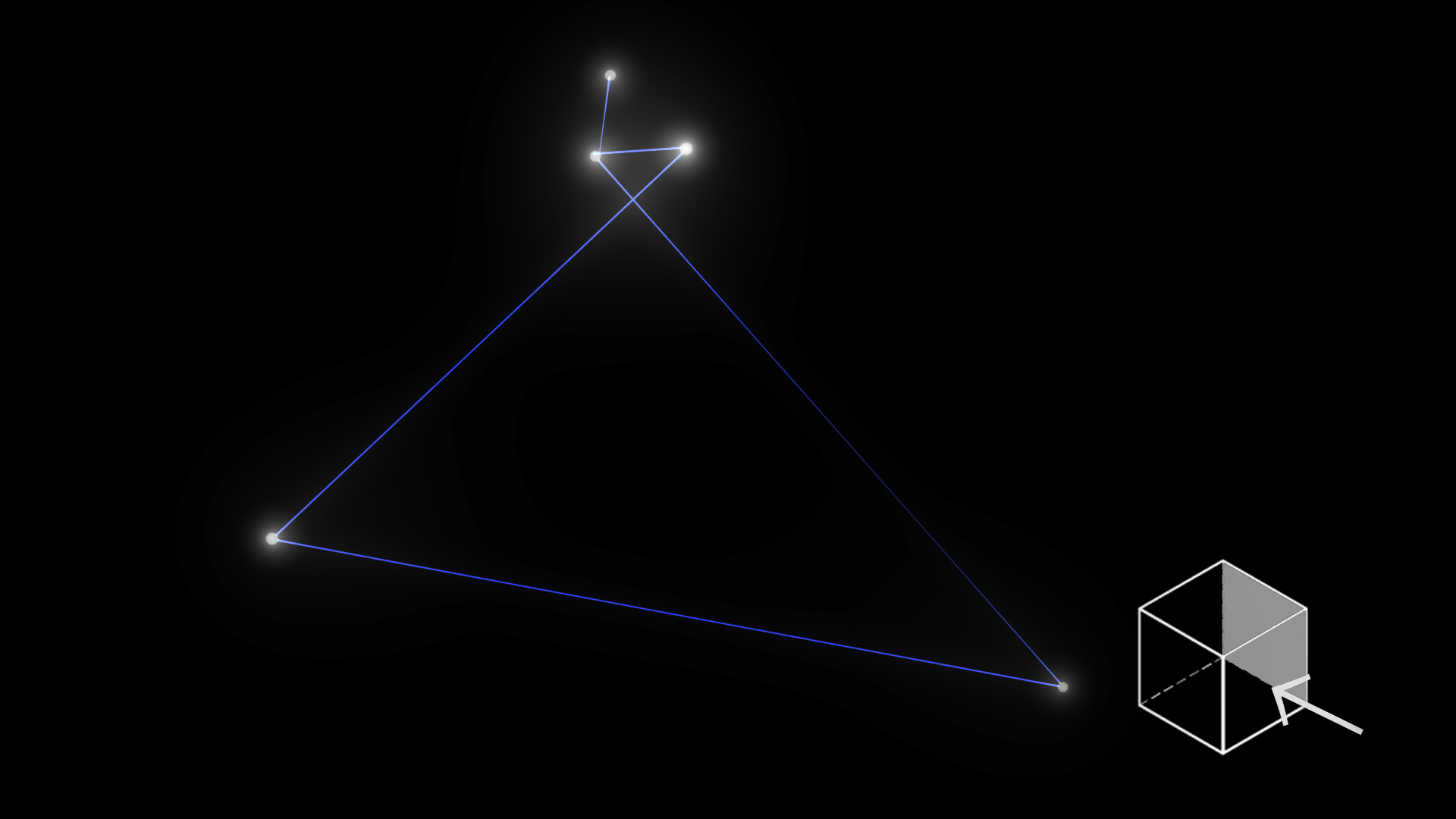}\\
Figure 18. Libra-Left view. Author: Laserna, H. (2016).\\

\subsection{Scorpio constellation}
\includegraphics[scale=0.1]{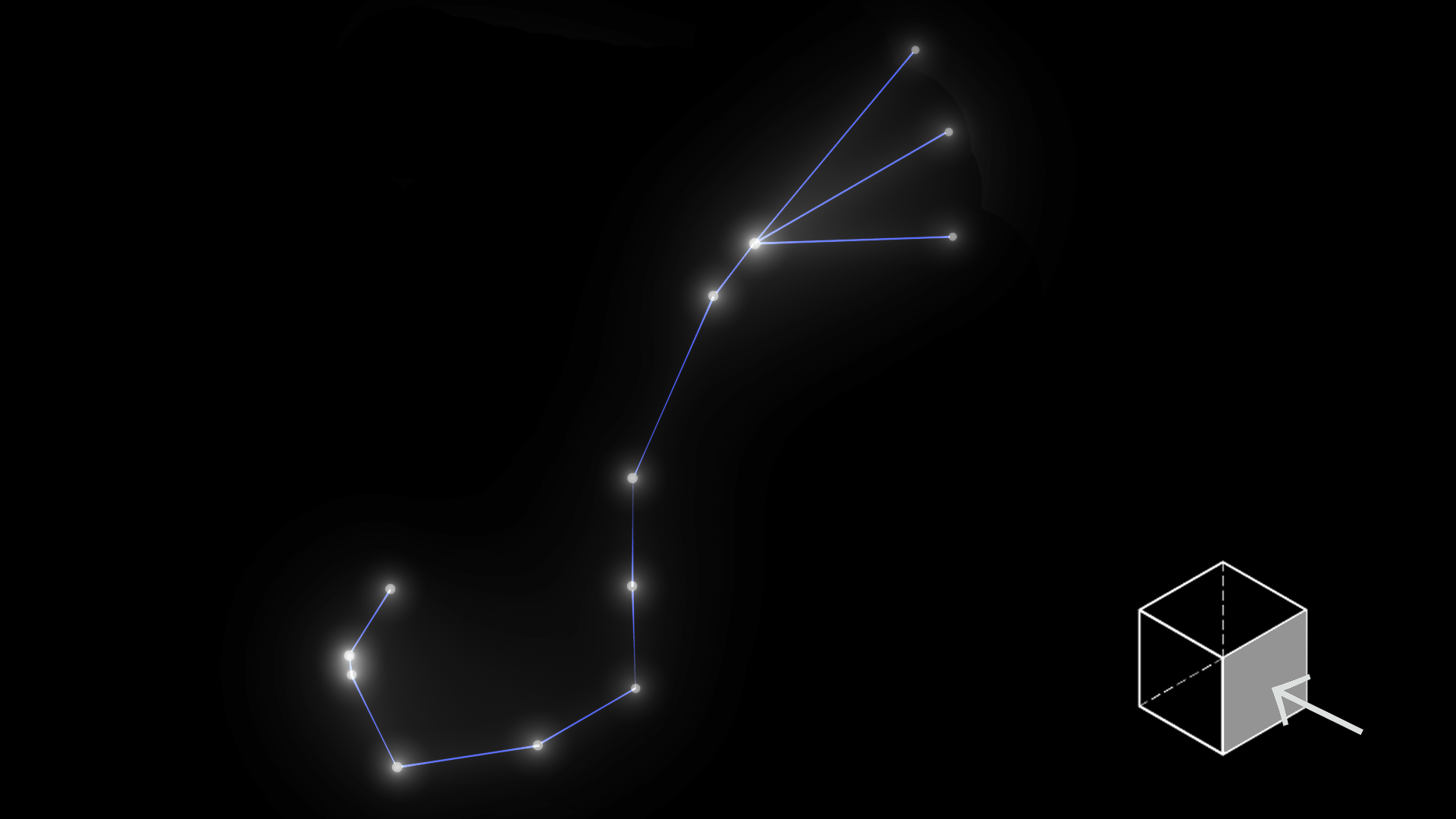}\\
Figure 19. Scorpio-Front view. Author: Laserna, H. (2016).\\

\includegraphics[scale=0.1]{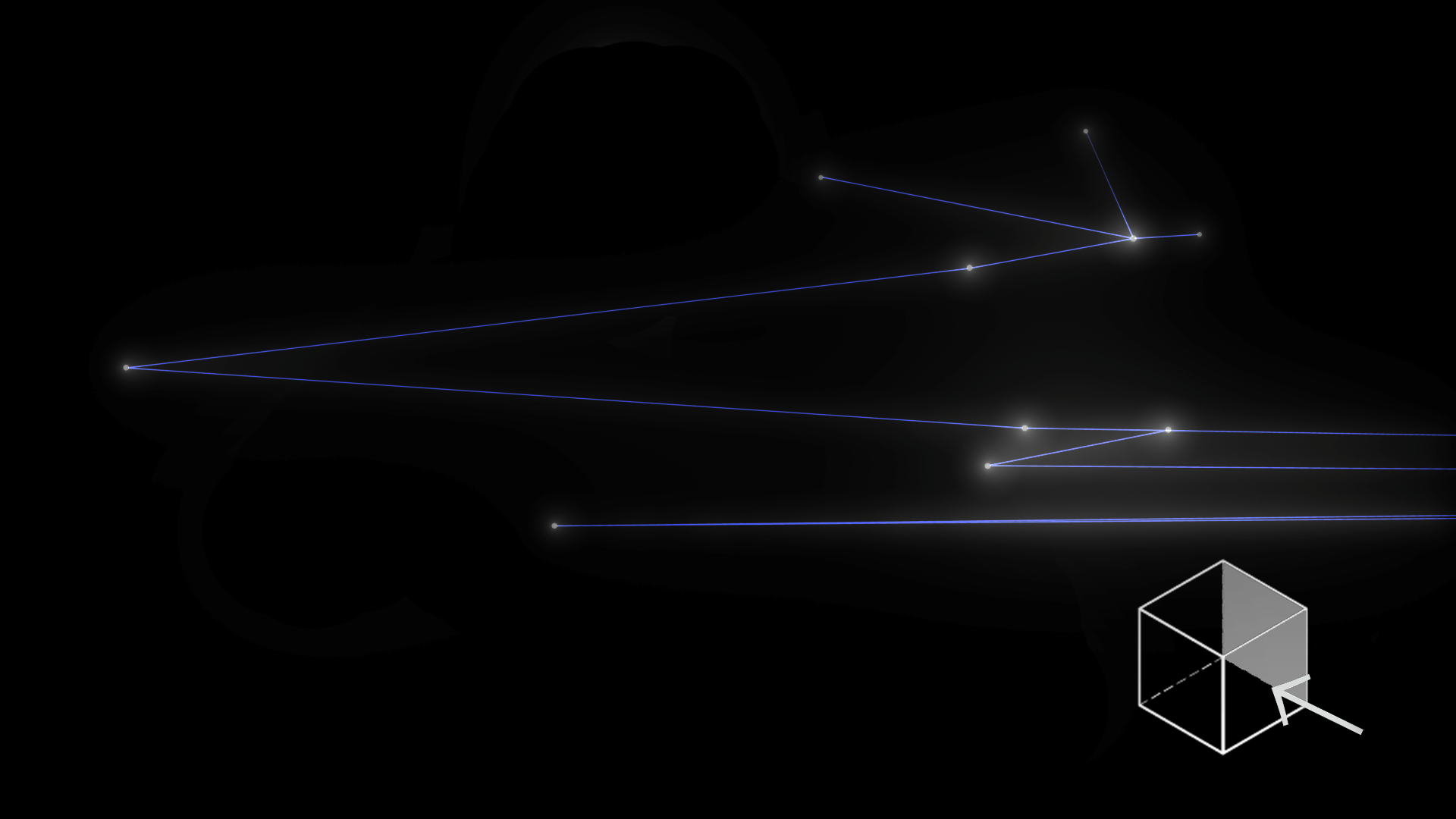}\\
Figure 20. Scorpio-Left view. Author: Laserna, H. (2016).\\

\subsection{Ophiuchus constellation}
\includegraphics[scale=0.1]{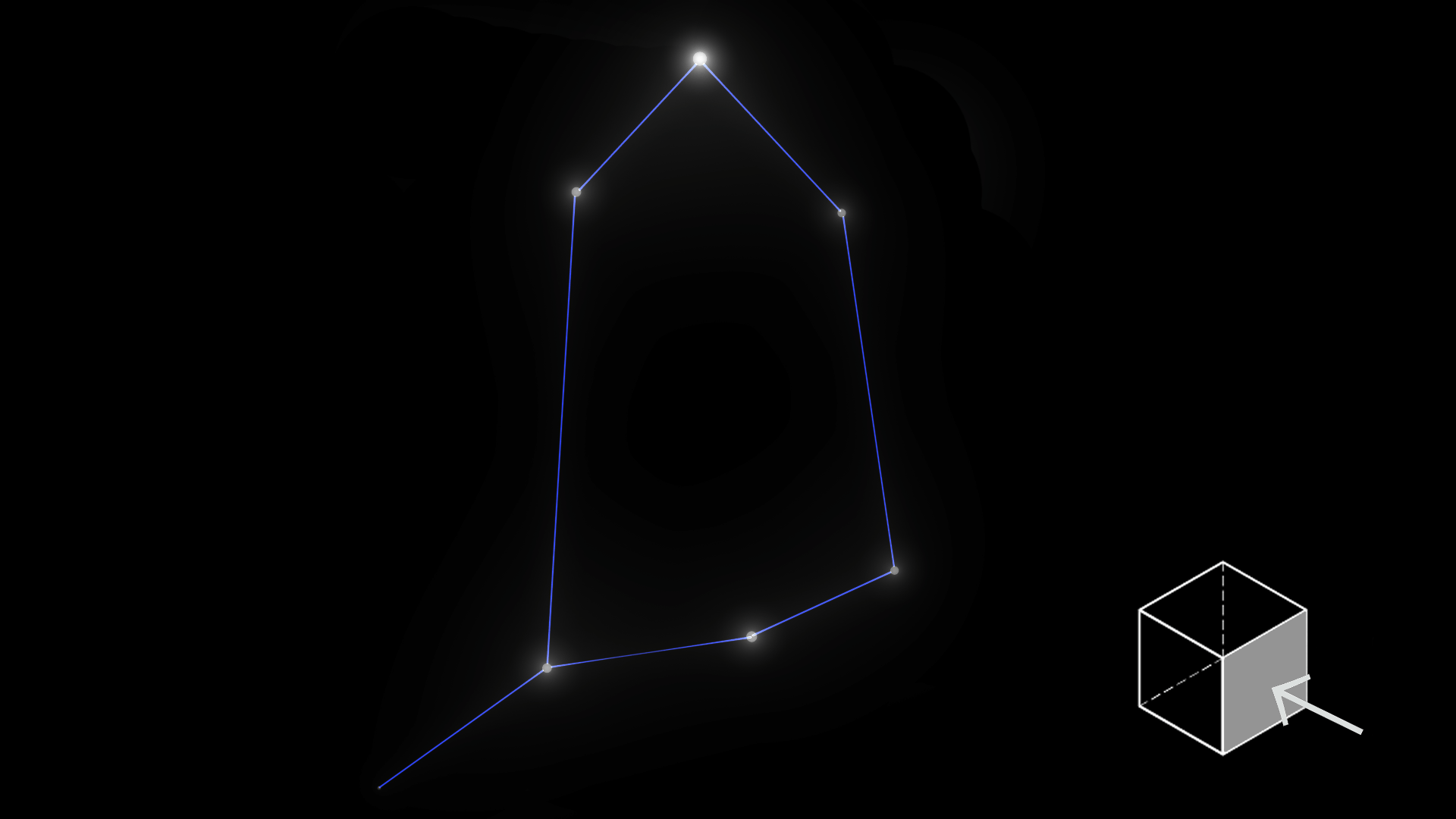}\\
Figure 21. Ophiuchus-Front view. Author: Laserna, H. (2016).\\

\includegraphics[scale=0.1]{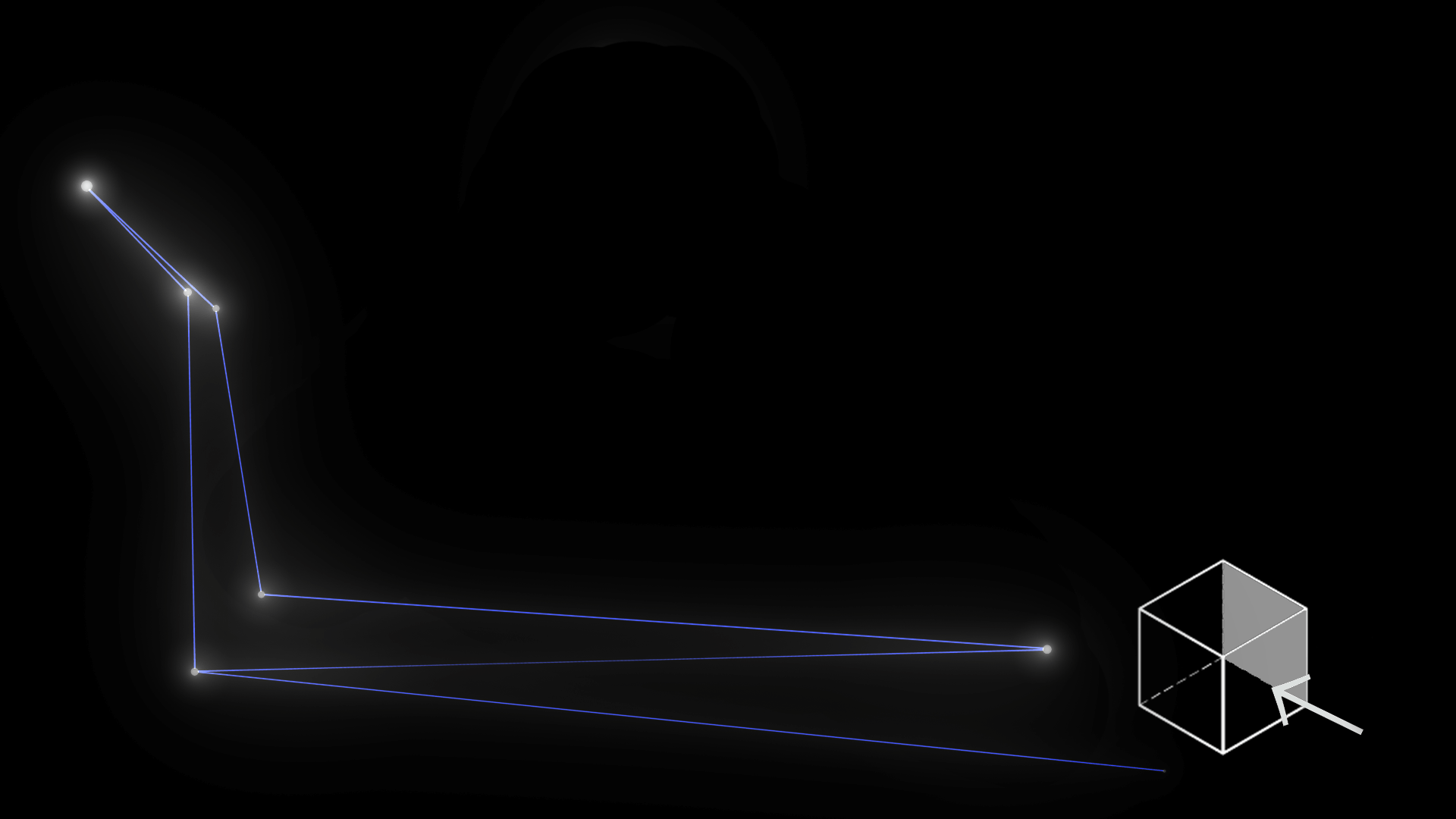}\\
Figure 22. Ophiuchus-Left view. Author: Laserna, H. (2016).\\

\subsection{Sagittarius constellation}
\includegraphics[scale=0.1]{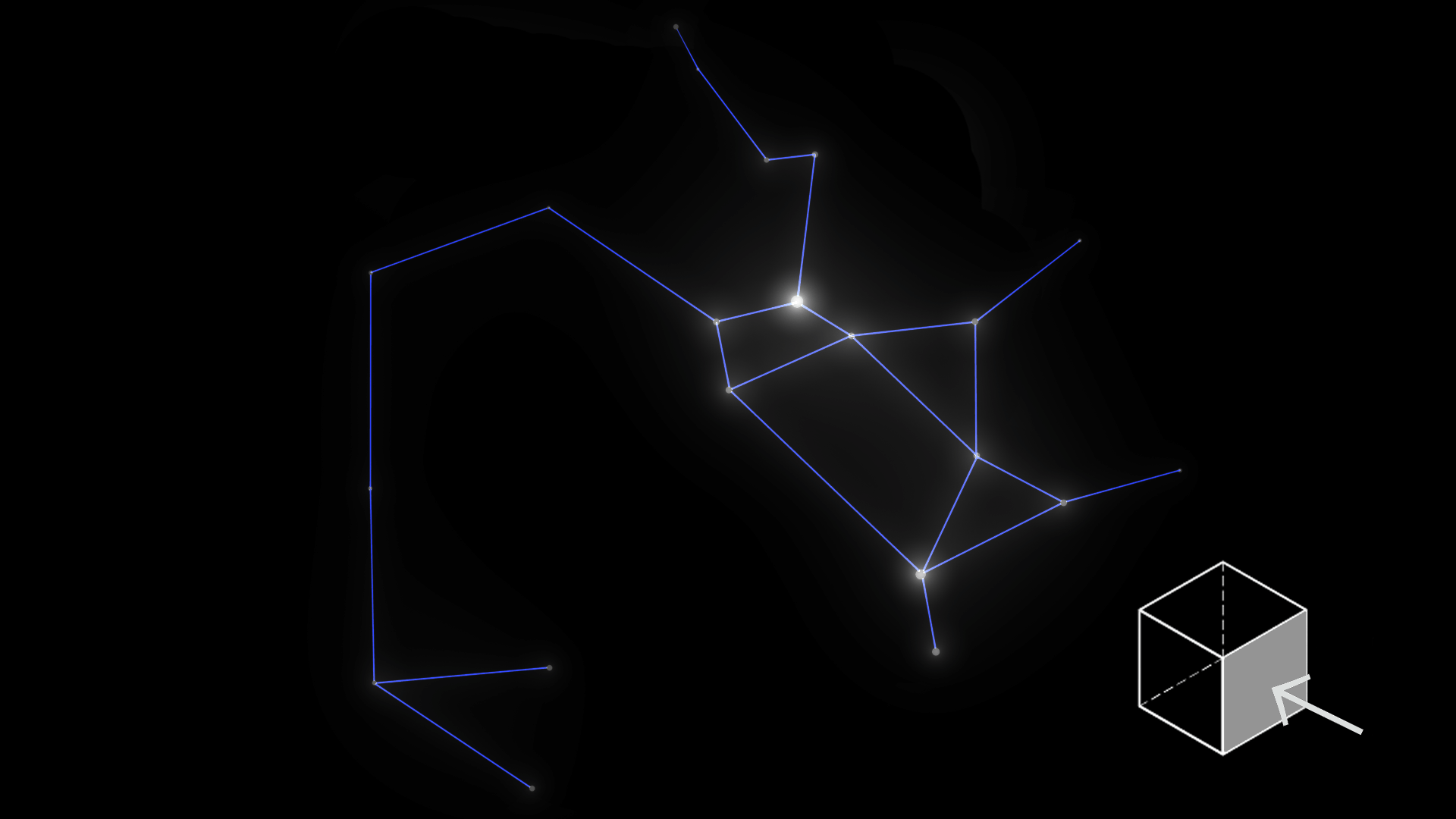}\\
Figure 23. Sagittarius-Front view. Author: Laserna, H. (2016).\\

\includegraphics[scale=0.1]{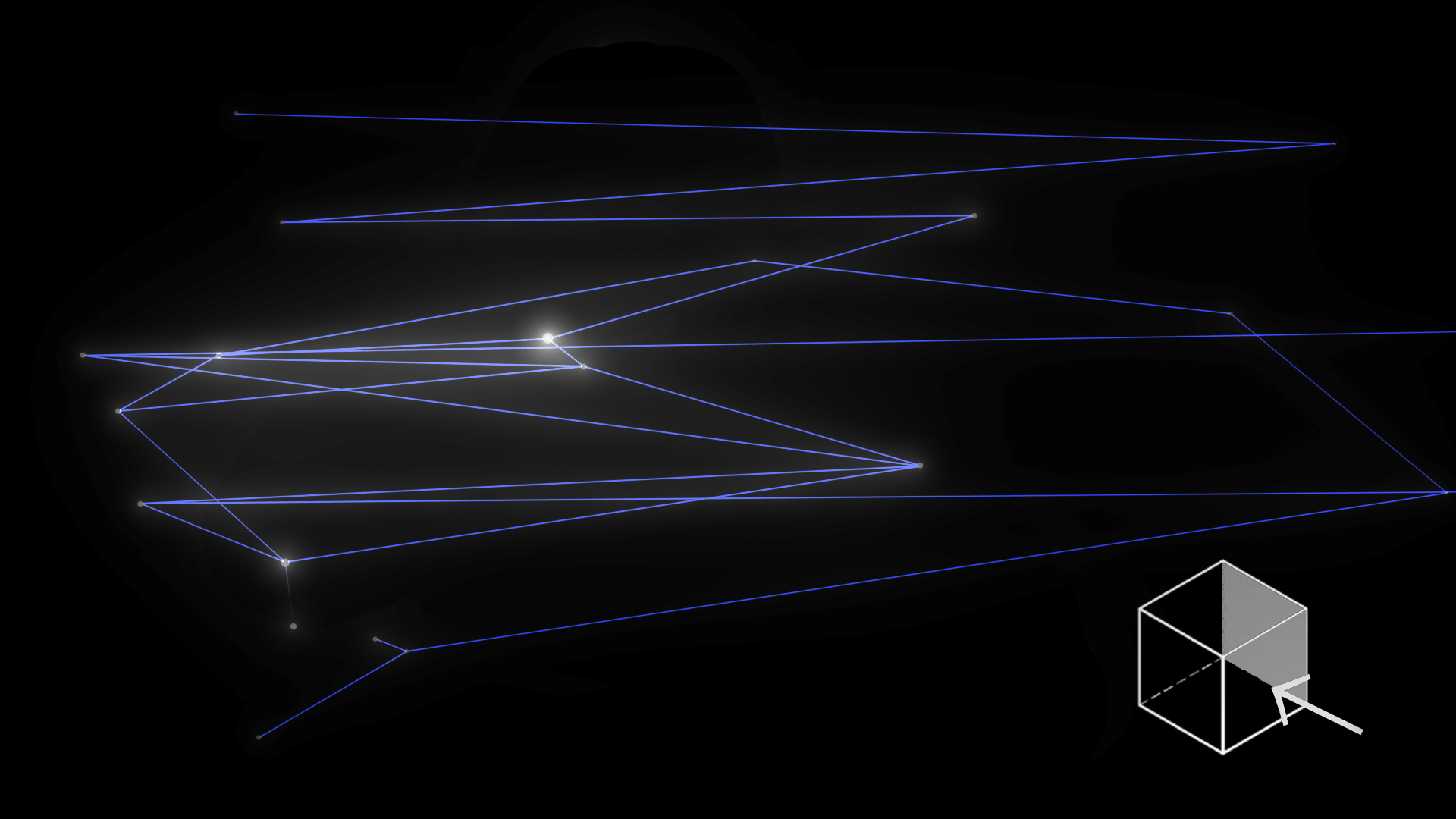}\\
Figure 24. Sagittarius-Left view. Author: Laserna, H. (2016).\\

\subsection{Capricorn constellation}
\includegraphics[scale=0.1]{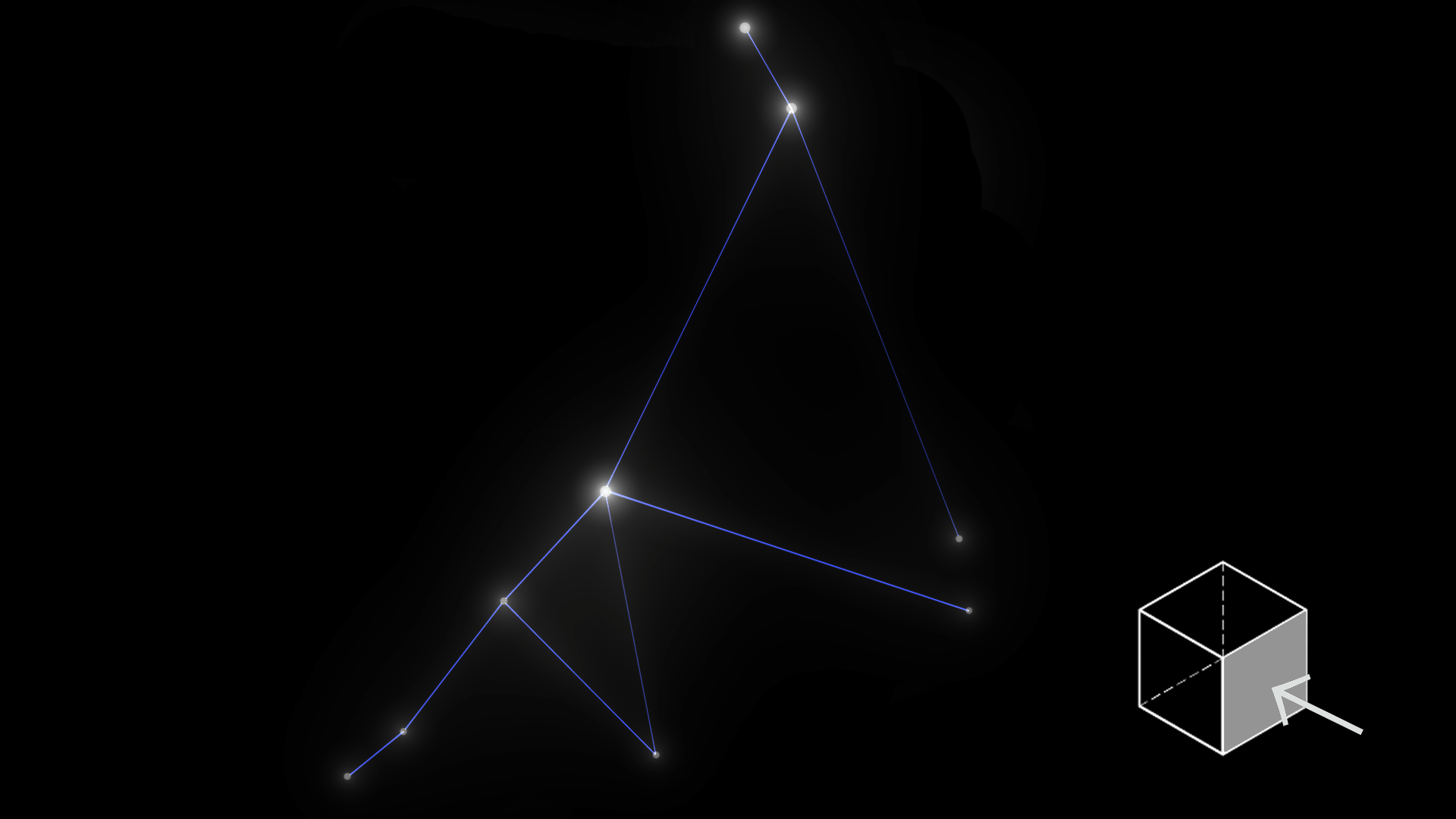}\\
Figure 25. Capricorn-Front view. Author: Laserna, H. (2016).\\

\includegraphics[scale=0.1]{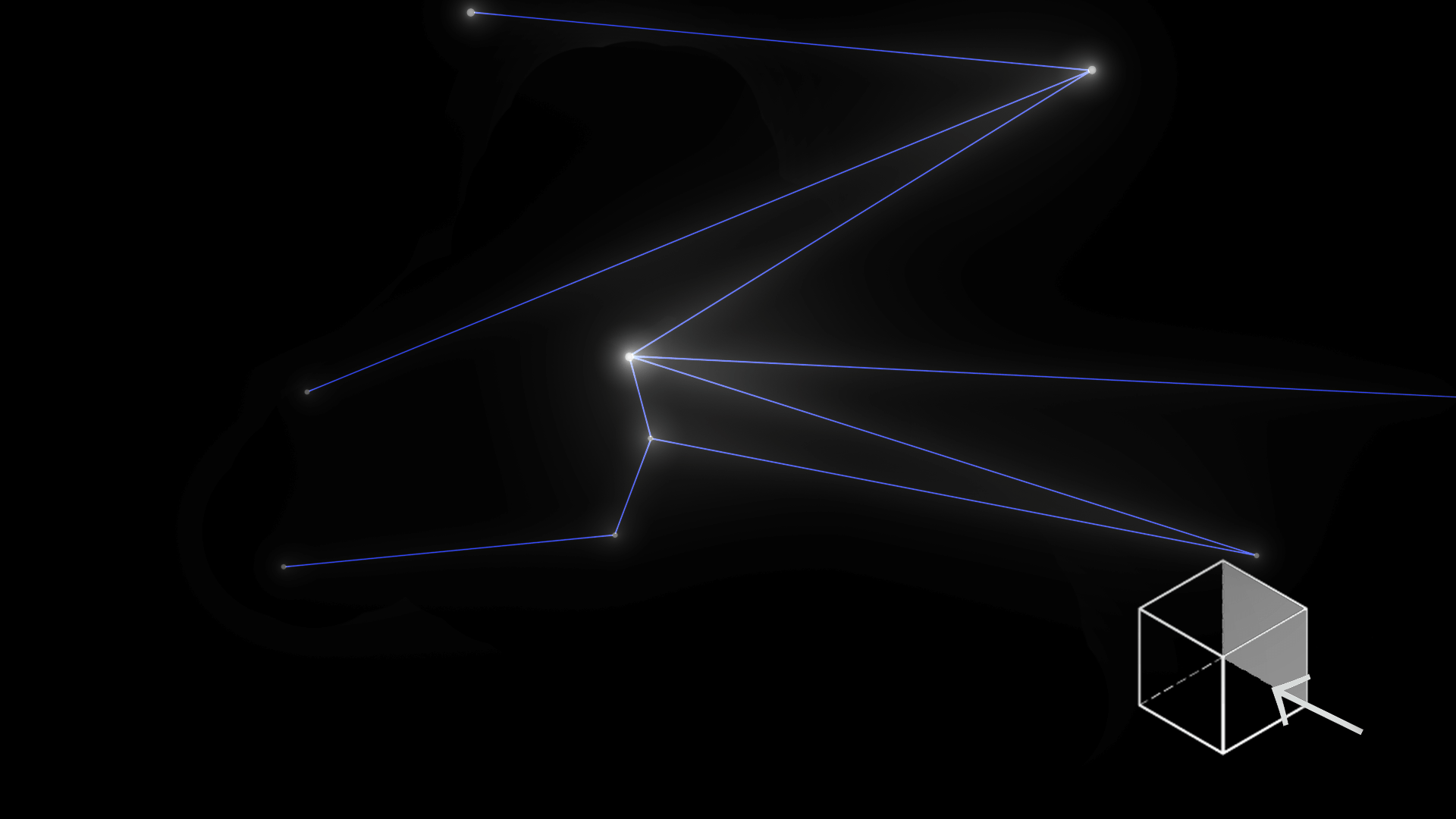}\\
Figure 26. Capricorn-Left view. Author: Laserna, H. (2016).\\

\subsection{Aquarius constellation}
\includegraphics[scale=0.1]{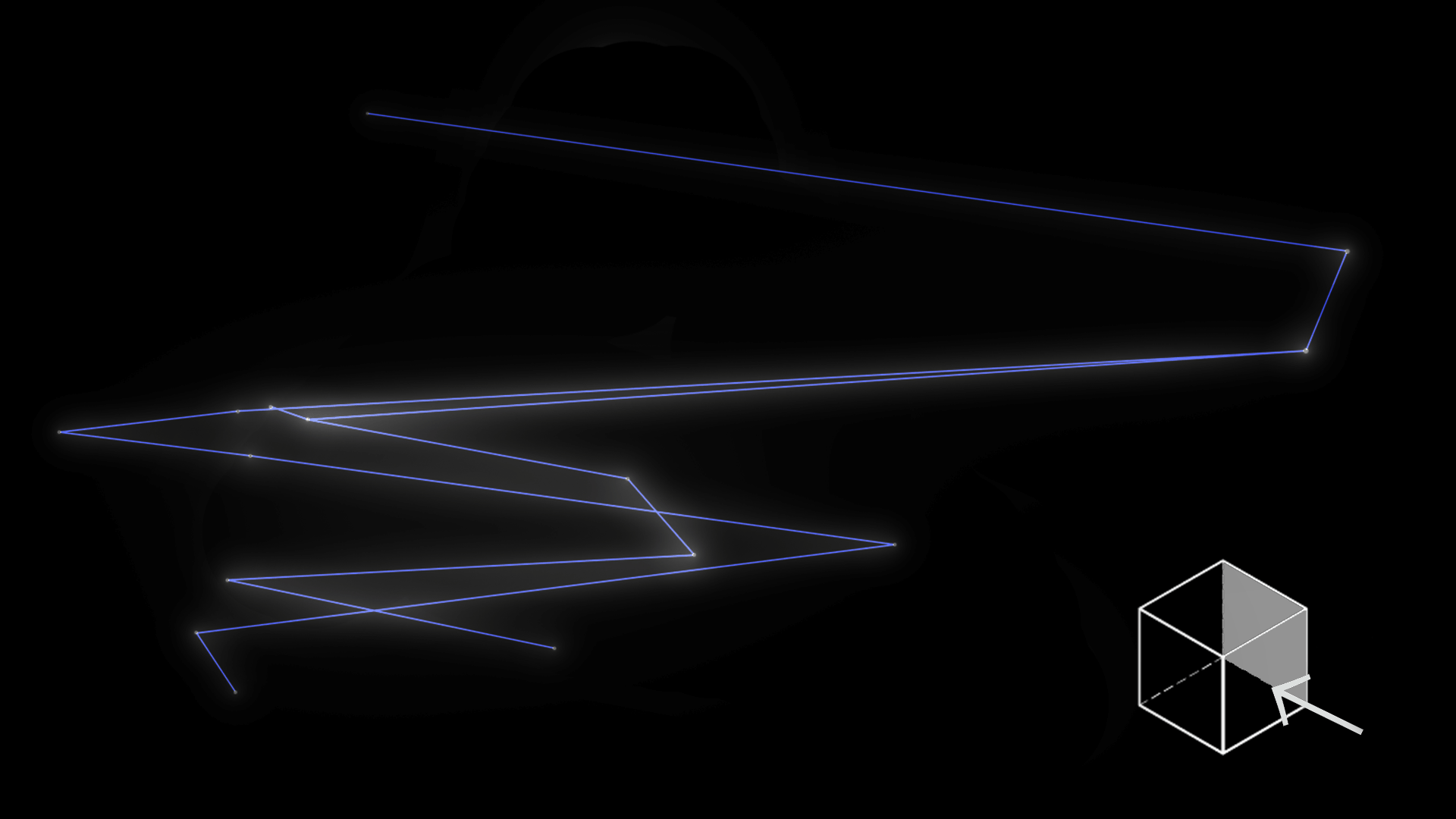}\\
Figure 27. Aquarius-Front view. Author: Laserna, H. (2016).\\

\includegraphics[scale=0.1]{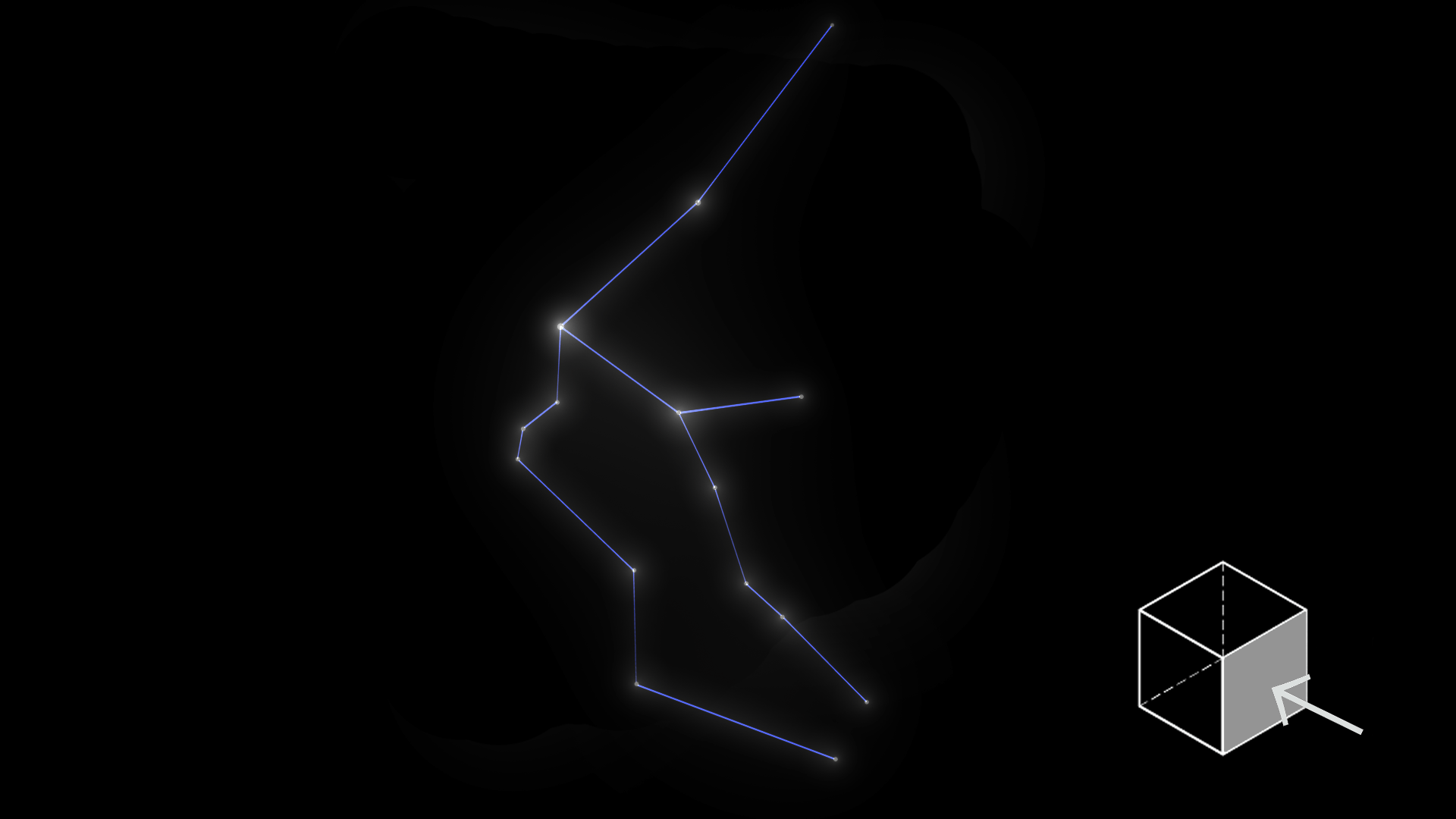}\\
Figure 28. Aquarius-Left view. Author: Laserna, H. (2016).\\

\section{Conclusions}
The use of computational tools in the teaching of physics is a didactic strategy that allows improving the processes of formalization and conceptualization of physical models in astronomy.\\
\\
In the teaching-learning processes of physics, it is important to implement different methodologies that strengthen comprehension processes, as in the case of simulations, since interactive tools allow us to demonstrate the phenomenology involved in physical models.\\
\\
Simulations in the field of astronomy make it possible to demonstrate the great distances between the celestial bodies through the different changes of scale that can be obtained in the use of computational methods.\\
\\
The analysis of the celestial bodies through the different frames of reference allowed to demonstrate that the concept of stellar structures or systems are intrinsically related to the position of the observers.
The star systems or structures obey cultural constructions that human beings have defined as cosmogonies that have subsequently been mathematized and have become a scientific explanation of the universe.

\section{References}
\begin{itemize}
\item Portilla, G. (2012). Elementos De Astronomía De Posición. UNIBIBLOS. Colombia.
\item Sistemas de proyecciones ortogonales en dibujo técnico y vistas. (s.f.). Recuperado de: www.larapedia.com/ingenieria$\_y\_tecnologia/sistemas$ $\_de\_proyecciones\_ortogonales$ $\_en\_dibujo\_tecnico\$_y$ $\_vistas.html$
\item Karttunen, H. et al. (1996). Fundamental Astronomy. Springer-Velrag. Heidelberg. ed. Portilla, G. (2001). Astronomía para Todos. UNIBIBLOS. Colombia.
\item Ronan, C. A. Los amantes de la astronomía. Editorial Blume. Barcelona.
\item Vives, T. J. (1971). Astronomía de posición. Alhambra. Madrid.
\item Toomer, G. J. (1998). Ptolemy’s Almagest. Princenton University Press. Princenton.
\end{itemize}
\end{multicols}
\end{document}